\documentclass[a4wide,12pt,onecolumn,oneside,notitlepage,final]{article}
\usepackage{amsmath}
\usepackage{a4wide}
\usepackage{amsfonts}
\usepackage{amstext}
\usepackage{amssymb}
\usepackage{amsthm}
\usepackage{graphics}
\usepackage{graphicx}
\usepackage{color}
\usepackage{verbatim}
\usepackage{epic,eepic}

\newcommand{\calS}{\mathcal{S}}

\newtheorem{theorem}{Theorem}[section] 

\newtheorem{proposition}{Proposition}[section]

\def\D{\Delta}
\def\l{u}

\title{{\bf Boxed Skew Plane Partition and \\ Integrable Phase Model}}
\author{Keiichi Shigechi
{\footnote {\tt E-mail: shigechi@monet.phys.s.u-tokyo.ac.jp}}\,\,\, and
\setcounter{footnote}{2}
Masaru Uchiyama
{\footnote {\tt E-mail: uchiyama@monet.phys.s.u-tokyo.ac.jp }}\, \\ \\
{\it Department of Physics, Graduate School of Science,}\\
{\it University of Tokyo,}\\
{\it Hongo 7-3-1, Bunkyo-ku, Tokyo 113-0033, Japan}\\
}
\date{}

\begin{document}
\maketitle

\begin{abstract}
We study the relation between the boxed skew plane partition and 
the integrable phase model. We introduce a generalization of a scalar 
product of the phase model and calculate it in two ways; the first one   
in terms of the skew Schur functions, and another one by use of 
the commutation relations of operators. In both cases, a generalized scalar 
product is expressed as a determinant. We show that a special choice of 
the spectral parameters of a generalized scalar product gives 
the generating function of the boxed skew plane partition. 
\end{abstract}


short title: Boxed Skew Plane Partition and Integrable Phase Model 

\baselineskip=16pt

\newpage

\section{Introduction}
Interplay between statistical mechanics and combinatorics is one of 
the most interesting features in mathematical physics \cite{Bax}. 
Recently, a connection between the integrable phase model and enumeration of 
plane partition  was shown in \cite{Bog}. 
The integrable phase model is a special limit of the $q$-boson model which is 
also related to the XXZ model \cite{qboson1, phase, qboson2}, 
and it can be solved applying the Quantum Inverse 
Scattering Method (QISM) \cite{F, KorBogIze}.

Study of generating function of plane partition \cite{Andrews, Stanley, Mac}
has attracted many researchers going back to P.~A.~MacMahon. 
For an illuminating story of plane partition involving 
Alternating Sign Matrix theorems and the six-vertex model with 
domain wall boundary condition \cite{Ku, Zeil}, 
we refer to \cite{Br}.
Recently, correlation functions for random (skew) plane partition were studied 
in a more general setting in terms of the free-fermion fields \cite{OkRe1, OkRe2}. 
There the generating function of unrestricted plane partition is 
presented as just a normalization constant, or a zero-point function. 

In this article, we generalize \cite{Bog} and argue the connection between 
the integrable phase model and the boxed skew plane partition. 
The generating functions of the boxed skew plane partition are given by 
a special parameter choice of the generalized scalar products of the phase model. 
We calculate the generalized scalar products with generic parameters 
in two ways. Namely, the one is expressed in terms of
the Schur function; the eigenfunction for the phase model, and 
another is, through the relations from QISM, 
expressed in terms of a determinant of matrix with elements of 
two-point functions and one-point functions. 

The paper is organized as follows. 
In Section \ref{phase}, we introduce the phase model as a quantum integrable system. 
The integrability of this model is understood in the context of the QISM. 
Through graphical interpretations, 
the action of the monodromy operators on a Fock state turns out to be written 
in terms of the skew Schur functions. 
A scalar product, which is defined as the inner product of 
the $N$-particle states of the phase model, is generalized. 
The inhomogeneous phase model is also considered. 
In Section \ref{sec:PP}, we briefly introduce the skew plane partition in a box 
and its generating function.
In Section \ref{SPP-GSP}, we see the connection between a generalized scalar product 
of the phase model and the skew plane partition. 
A sequence of operators in a generalized scalar product 
corresponds to the shape of the skew 
plane partition, and the number of lattice sites is equal to the height. 
The special choice of the spectral parameters naturally 
gives the generating function of the skew plane partition 
through graphical observations. 
A generalized scalar product is calculated in two ways 
in Section \ref{skewschur} and Section \ref{GSP-2}. The first 
one is by the help of the expression in the skew Schur functions. 
Another one is an inductive application of 
the commutation relations of the  monodromy operators. 
In Section \ref{genefuncbpp}, we show that the generating function of 
the boxed plane partition has a determinant expression, 
which is a by-product of the explicit expression of a generalized scalar product
obtained in Section \ref{GSP-2}. 
Section \ref{concl} is devoted to the conclusion. 

%%%%% Phase-Model
\section{The Integrable Phase Model}
\label{phase}
In this section, we introduce the integrable phase model and review the results on 
this model concerning its relation to the skew Schur functions based on~\cite{Bog}.
We also introduce and generalize the scalar product of the wave functions of the 
phase model. This is the main quantity which is viewed as the 
generating function of the plane partition in the subsequent sections. 

\subsection{Integrability of phase model}\label{phase-YBR}
We introduce the phase model as an integrable system. 
Consider a one-dimensional lattice of length $M+1$ and label each site 
by $i=0,\cdots,M$. Let $\phi_{i},\phi_{i}^{\dagger}$ denote  
the operators satisfying the following commutation relations
\begin{eqnarray}
[N_{i},\phi_{j}]=-\phi_{i}\delta_{ij},\ [N_{i},\ 
\phi^{\dagger}_{j}]=\phi_{i}^{\dagger}\delta_{ij}, \
[\phi_{i},\phi^{\dagger}_{j}]=\pi_{i}\delta_{ij}
\end{eqnarray}
for $i,j=0,\cdots,M$
where $N_{i}$ is the number operator and 
$\pi_{i}$ is the vacuum projector. 
The $L$-operator for the phase model is defined as 
\begin{eqnarray}
L_{i}(u)=\left(
  \begin{array}{cc}
    u^{-1}   &  \phi^{\dagger}_{i}  \\
    \phi_{i}   &  u  \\
  \end{array}
\right)
\end{eqnarray}
with the spectral parameter $u\in \mathbb{R}$. 
The local Fock states at site $j$ are created from the 
local vacuum state $|0\rangle_j$ such as 
$|n_j\rangle=(\phi_j^{\dagger})^{n_j}|0\rangle_j, 
N_j|n_j\rangle_j=n_j|n_j\rangle_j$. The states satisfy
\begin{eqnarray}
\begin{aligned}
&\phi_{j}|0\rangle_{j}=0,\\
&\phi_{j}|n_{j}\rangle_{j}=|n_{j}-1\rangle_j,\ \ \ 
\phi_{j}^{\dagger}|n_{j}\rangle_{j}=|n_{j}+1\rangle_{j}, \\
&N_{j}|n_{j}\rangle_{j}=n_{j}|n_{j}\rangle_{j}. 
\end{aligned}
\end{eqnarray}
The inner product of two Fock states is normalized, 
$_{i}\langle m|n\rangle_{j}=\delta_{ij}\delta_{nm}$. 

Let us introduce the $R$-matrix (a $4\times 4$ matrix) of the form
\begin{eqnarray}
\begin{aligned}
&R_{11}(u,v)=R_{44}(u,v)=f(v,u), \\
&R_{22}(u,v)=R_{33}(u,v)=g(v,u), \\
&R_{23}(u,v)=1,
\end{aligned}
\end{eqnarray}
where 
\begin{eqnarray}
f(v,u)=\frac{u^{2}}{u^{2}-v^{2}},\ \ g(v,u)=\frac{uv}{u^{2}-v^{2}},
\label{fg}
\end{eqnarray}
and the other elements are zero. Then we can find that the $L$-operator satisfies 
the intertwining relation
\begin{eqnarray}
R(u,v)(L_{n}(u)\otimes L_{n}(v))=(L_{n}(v)\otimes L_{n}(u))R(u,v).
\label{YBLR}
\end{eqnarray}
The monodromy matrix is defined as 
\begin{eqnarray}
T(u)=L_{M}L_{M-1}\cdots L_{0}=\left(
  \begin{array}{cc}
    A(u)   & B(u)   \\
    C(u)   & D(u)   \\
  \end{array}
\right).\label{MonoMatrix}
\end{eqnarray}
Hereafter, we refer to the elements of the monodromy matrix as the monodromy operators.
The commutation relations of the monodromy operators can be obtained from
\begin{eqnarray}
R(u,v)(T(u)\otimes T(v))=(T(v)\otimes T(u))R(u,v).
\label{YBTR}
\end{eqnarray}
Several explicit relations are listed in the following. 
\begin{eqnarray}
&&[A(u),A(v)]=[B(u),B(v)]=[C(u),C(v)]=[D(u),D(v)]=0, \label{AA}\\
%&&C(u)B(v)=C(v)B(u), \\
%%%%%%&&C(u)A(v)=\frac{u}{v}C(v)A(u),\ C(u)D(v)=\frac{v}{u}C(v)D(u),\\
&&g(v,u)C(u)A(v)=f(v,u)C(v)A(u), \label{CA1}\\
%&&f(v,u)C(u)D(v)=g(v,u)C(v)D(u), \\
%%%%%%%&&A(u)B(v)=\frac{v}{u}A(v)B(u),\ D(u)B(v)=\frac{u}{v}D(v)B(u),\\
%&&f(v,u)A(u)B(v)=g(v,u)A(v)B(u), \\
&&g(v,u)D(u)B(v)=f(v,u)D(v)B(u), \label{DB1}\\
&&C(u)B(v)=g(u,v)\{A(u)D(v)-A(v)D(u)\}=g(u,v)\{D(u)A(v)-D(v)A(u)\},\\
&&C(u)A(v)=f(v,u)A(v)C(u)+g(u,v)A(u)C(v), \label{CA2}\\
&&D(u)B(v)=f(v,u)B(v)D(u)+g(u,v)B(u)D(v), \label{DB2}
%&&A(u)B(v)=f(u,v)B(v)A(u)-g(u,v)B(u)A(v), \\
%&&C(u)D(v)=f(u,v)D(v)C(u)-g(u,v)D(u)C(v), \\
%&&[D(u),A(v)]=g(u,v)\{B(u)C(v)-B(v)C(u)\}.
\end{eqnarray}
Note that these commutation relations can be reduced from those of 
the $q$-boson model. 

\subsection{Graphical representation of operators}\label{graphical}
The monodromy operators $A,B,C$ and $D$ are expressed as the linear combination of 
products of elements of $L$-operators. For later
convenience, we introduce a graphical representation of 
the $L$-operator and the monodromy matrix in this subsection. 

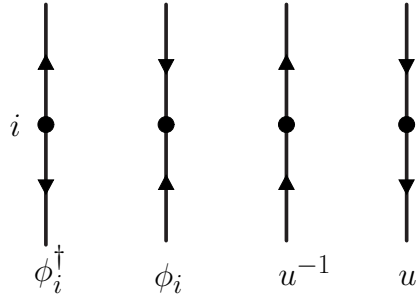
\begin{figure}[htbp]
\begin{center}  %%  fig-1.tex
{\setlength{\unitlength}{0.00083333in}  
\begingroup\makeatletter\ifx\SetFigFont\undefined%
\gdef\SetFigFont#1#2#3#4#5{%
  \reset@font\fontsize{#1}{#2pt}%
  \fontfamily{#3}\fontseries{#4}\fontshape{#5}%
  \selectfont}%
\fi\endgroup%
{\renewcommand{\dashlinestretch}{30}
\scalebox{0.5}{
\begin{picture}(5069,4305)(0,-10)
\allinethickness{0.1cm}
\path(450,600)(450,3600)
\path(1950,650)(1950,3600)
\path(3450,650)(3450,3600)
\path(4950,650)(4950,3600)
\thinlines
\put(450,2100){\blacken\ellipse{200}{200}}
\put(1950,2100){\blacken\ellipse{200}{200}}
\put(3450,2100){\blacken\ellipse{200}{200}}
\put(4950,2100){\blacken\ellipse{200}{200}}
\blacken\path(450,1230)(355,1410)(545,1410)(450,1230)  
\blacken\path(1950,1470)(1855.000,1290)(2045.000,1290)(1950.000,1470)
\blacken\path(3450,1470)(3355.000,1290)(3545.000,1290)(3450,1470)
\blacken\path(4950,1230)(4855,1410)(5045.000,1410)(4950,1230)
\blacken\path(450,2970)(355,2790)(545,2790)(450,2970)
\blacken\path(1950,2730)(1855,2910)(2045,2910)(1950,2730)
\blacken\path(3450,2970)(3355,2790)(3545,2790)(3450,2970)
\blacken\path(4950,2730)(4855,2910)(5045,2910)(4950,2730)
\put(0,1950){\makebox(0,0)[lb]{{\SetFigFont{29}{34.8}{\rmdefault}{\mddefault}{\updefault}$i$}}}
\put(300,0){\makebox(0,0)[lb]{{\SetFigFont{29}{34.8}{\rmdefault}{\mddefault}{\updefault}$\phi^{\dagger}_{i}$}}}
\put(1800,0){\makebox(0,0)[lb]{{\SetFigFont{29}{34.8}{\rmdefault}{\mddefault}{\updefault}$\phi_{i}$}}}
\put(4850,100){\makebox(0,0)[lb]{{\SetFigFont{29}{34.8}{\rmdefault}{\mddefault}{\updefault}$u$}}}
\put(3350,100){\makebox(0,0)[lb]{{\SetFigFont{29}{34.8}{\rmdefault}{\mddefault}{\updefault}$u^{-1}$}}}
\end{picture}
}}}
\end{center}
  \caption{Vertex representation for the elements of $L$-operator.}
  \label{fig1}
\end{figure}

We assign a vertex with two vertical arrowed edges to each element of 
$L$-operators. Each arrow is directed upward or downward. 
The vertex is indexed by $i$ for the $i$-th site. 
Pairs of arrows stand for $\phi_i, \phi_i^\dagger, u$ and $u^{-1}$, 
respectively (see Figure  \ref{fig1}). 
% figure

This vertex representation is useful for a calculation of products of $L$-operators. 
We attach vertices vertically by the edge when the direction 
of the arrow between two neighboring vertices are the same. 
This gives a one-dimensional vertical graph. 
Then, each element of a product of $L$-operators can be represented as 
summation over graphs of all arrow configurations 
with certain fixed boundaries. 
For instance, the $(1,1)$-element of $L_{i}(u)L_{j}(u)$ is 
$u^{-2}+\phi_{i}^{\dagger}\phi_{j}$ and the corresponding graphs have 
both boundary arrows upward. 

The monodromy operators $A,B,C$ and $D$ have also 
one-dimensional graph representations. From their 
definitions (\ref{MonoMatrix}), each operator can be represented as 
a sum over graphs of all possible arrow configurations on a vertical lattice with $M+1$ 
vertices and fixed boundary arrows. 
Here the lattice sites are numbered by $i=0,\cdots,M$ from the bottom. 
Operator $B(u)$ (resp. $C(u)$) has the top 
and bottom arrows pointing outward (resp. inward). 
Operator $A(u)$ (resp. $D(u)$) has the top and bottom arrows 
pointing upward (resp. downward). 

The number of $\phi^{\dagger}$ is bigger (resp. less) by one 
than that of $\phi$ in every possible arrow configuration 
of $B(u)$ (resp. $C(u)$). In this sense, the operators $B(u)$ and $C(u)$ 
have the property of the creation and annihilation operator, respectively. 
On the other hand, the number of $\phi^{\dagger}$ is exactly equal to 
that of $\phi$ in every possible arrow configurations of $A(u)$ and $D(u)$.

\begin{figure}[htbp]
  \begin{center}
{\setlength{\unitlength}{0.00083333in}
\begingroup\makeatletter\ifx\SetFigFont\undefined%
\gdef\SetFigFont#1#2#3#4#5{%
  \reset@font\fontsize{#1}{#2pt}%
  \fontfamily{#3}\fontseries{#4}\fontshape{#5}%
  \selectfont}%
\fi\endgroup%
{\renewcommand{\dashlinestretch}{30}
\scalebox{0.5}{
\begin{picture}(4915,5159)(0,-10)
\put(0,4050){\makebox(0,0)[lb]{{\SetFigFont{20}{34.8}{\rmdefault}{\mddefault}{\updefault}$M$}}}
\put(0,3200){\makebox(0,0)[lb]{{\SetFigFont{18}{34.8}{\rmdefault}{\mddefault}{\updefault}$M-1$}}}
\put(4725,3300){\blacken\ellipse{120}{120}}
\put(4725,3300){\ellipse{120}{120}}
\put(4725,1500){\blacken\ellipse{120}{120}}
\put(4725,1500){\ellipse{120}{120}}
\put(4725,2400){\blacken\ellipse{120}{120}}
\put(4725,2400){\ellipse{120}{120}}
\put(3525,4200){\blacken\ellipse{120}{120}}
\put(3525,4200){\ellipse{120}{120}}
\put(3525,3300){\blacken\ellipse{120}{120}}
\put(3525,3300){\ellipse{120}{120}}
\put(3525,1500){\blacken\ellipse{120}{120}}
\put(3525,1500){\ellipse{120}{120}}
\put(3525,2400){\blacken\ellipse{120}{120}}
\put(3525,2400){\ellipse{120}{120}}
\put(2325,4200){\blacken\ellipse{120}{120}}
\put(2325,4200){\ellipse{120}{120}}
\put(2325,3300){\blacken\ellipse{120}{120}}
\put(2325,3300){\ellipse{120}{120}}
\put(2325,1500){\blacken\ellipse{120}{120}}
\put(2325,1500){\ellipse{120}{120}}
\put(2325,2400){\blacken\ellipse{120}{120}}
\put(2325,2400){\ellipse{120}{120}}
\put(1125,4200){\blacken\ellipse{120}{120}}
\put(1125,4200){\ellipse{120}{120}}
\put(1125,3300){\blacken\ellipse{120}{120}}
\put(1125,3300){\ellipse{120}{120}}
\put(1125,1500){\blacken\ellipse{120}{120}}
\put(1125,1500){\ellipse{120}{120}}
\put(1125,2400){\blacken\ellipse{120}{120}}
\put(1125,2400){\ellipse{120}{120}}
\blacken\path(1125,4800)(975,4575)(1275,4575)(1125,4800)(1125,4800)
\path(1125,4800)(975,4575)(1275,4575)(1125,4800)(1125,4800)
\blacken\path(1125,3900)(975,3675)(1275,3675)(1125,3900)(1125,3900)
\path(1125,3900)(975,3675)(1275,3675)(1125,3900)(1125,3900)
\blacken\path(1125,2100)(975,1875)(1275,1875)(1125,2100)(1125,2100)
\path(1125,2100)(975,1875)(1275,1875)(1125,2100)(1125,2100)
\blacken\path(1125,1200)(975,975)(1275,975)(1125,1200)(1125,1200)
\path(1125,1200)(975,975)(1275,975)(1125,1200)(1125,1200)
\blacken\path(4725,900)(4875,1125)(4575,1125)(4725,900)(4725,900)
\path(4725,900)(4875,1125)(4575,1125)(4725,900)(4725,900)
\blacken\path(4725,1800)(4875,2025)(4575,2025)(4725,1800)(4725,1800)
\path(4725,1800)(4875,2025)(4575,2025)(4725,1800)(4725,1800)
\blacken\path(4725,3600)(4875,3825)(4575,3825)(4725,3600)(4725,3600)
\path(4725,3600)(4875,3825)(4575,3825)(4725,3600)(4725,3600)
\blacken\path(4725,4500)(4875,4725)(4575,4725)(4725,4500)(4725,4500)
\path(4725,4500)(4875,4725)(4575,4725)(4725,4500)(4725,4500)
\Thicklines
\path(4725,5100)(4725,3150)
\path(4725,2550)(4725,600)
\dashline{75.000}(4725,3150)(4725,2550)
\thinlines
\blacken\path(3525,1200)(3375,975)(3675,975)(3525,1200)(3525,1200)
\path(3525,1200)(3375,975)(3675,975)(3525,1200)(3525,1200)
\blacken\path(3525,4500)(3675,4725)(3375,4725)(3525,4500)(3525,4500)
\path(3525,4500)(3675,4725)(3375,4725)(3525,4500)(3525,4500)
\blacken\path(3525,3600)(3675,3825)(3375,3825)(3525,3600)(3525,3600)
\path(3525,3600)(3675,3825)(3375,3825)(3525,3600)(3525,3600)
\blacken\path(3525,1800)(3675,2025)(3375,2025)(3525,1800)(3525,1800)
\path(3525,1800)(3675,2025)(3375,2025)(3525,1800)(3525,1800)
\Thicklines
\path(3525,5100)(3525,3150)
\path(3525,2550)(3525,600)
\dashline{75.000}(3525,3150)(3525,2550)
\thinlines
\blacken\path(2325,900)(2475,1125)(2175,1125)(2325,900)(2325,900)
\path(2325,900)(2475,1125)(2175,1125)(2325,900)(2325,900)
\blacken\path(2325,4800)(2175,4575)(2475,4575)(2325,4800)(2325,4800)
\path(2325,4800)(2175,4575)(2475,4575)(2325,4800)(2325,4800)
\blacken\path(2325,3900)(2175,3675)(2475,3675)(2325,3900)(2325,3900)
\path(2325,3900)(2175,3675)(2475,3675)(2325,3900)(2325,3900)
\blacken\path(2325,2100)(2175,1875)(2475,1875)(2325,2100)(2325,2100)
\path(2325,2100)(2175,1875)(2475,1875)(2325,2100)(2325,2100)
\Thicklines
\path(2325,5100)(2325,3150)
\path(2325,2550)(2325,600)
\dashline{75.000}(2325,3150)(2325,2550)
\path(1125,5100)(1125,3150)
\path(1125,2550)(1125,600)
\dashline{75.000}(1125,3150)(1125,2550)
\put(3375,0){\makebox(0,0)[lb]{{\SetFigFont{34}{40.8}{\rmdefault}{\mddefault}{\updefault}$C$}}}
\put(2175,0){\makebox(0,0)[lb]{{\SetFigFont{34}{40.8}{\rmdefault}{\mddefault}{\updefault}$B$}}}
\put(900,0){\makebox(0,0)[lb]{{\SetFigFont{34}{40.8}{\rmdefault}{\mddefault}{\updefault}$A$}}}
\put(4500,0){\makebox(0,0)[lb]{{\SetFigFont{34}{40.8}{\rmdefault}{\mddefault}{\updefault}$D$}}}
\put(150,1350){\makebox(0,0)[lb]{{\SetFigFont{20}{34.8}{\rmdefault}{\mddefault}{\updefault}$0$}}}
\put(150,2250){\makebox(0,0)[lb]{{\SetFigFont{20}{34.8}{\rmdefault}{\mddefault}{\updefault}$1$}}}
\thinlines
\put(4725,4200){\blacken\ellipse{120}{120}}
\put(4725,4200){\ellipse{120}{120}}
\end{picture}
}}}
  \end{center}
  \caption{Basic arrow configurations of the operators $A, B, C$ and $D$.}
    \label{fig2}
\end{figure}
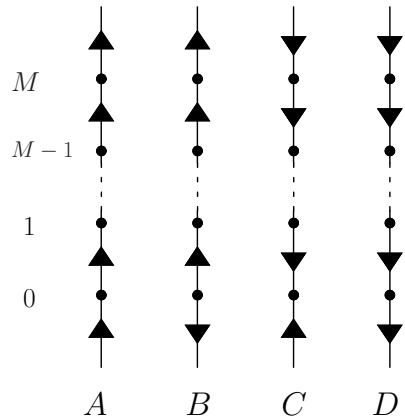

In later discussion, we call those shown in Figure \ref{fig2} the basic configurations. 
The other arrow configurations can be obtained by flipping some arrows upside-down.
% figure

\subsection{Fock states and partitions}\label{LFYdiagram}
In this subsection, we see the actions of the monodromy operators $A, B, C$ and $D$ 
on a Fock state. 
We introduce a bijection from an arrow configuration 
considered in the previous subsection to a lattice path configuration. 
By using this bijection, we express each matrix element of the operators 
with respect to the Fock states in terms of the skew Schur functions.

Any state of the phase model can be expanded by a linear combination of 
states of the form
\begin{eqnarray}
|\{n_{j}\}\rangle\equiv \prod_{j=0}^{M}|n_{j}\rangle_{j}, \qquad
N=\sum_{j=0}^{M}n_{j}, \qquad
n_{j}\ge0,
\end{eqnarray}
where $\{n_{j}\}$ stands for a configuration $\{n_{j}\}=(n_{0},\cdots,n_{M})$ 
with $n_{j}\ge 0$. 
We call this kind of state an $N$-particle basis. 
In particular, we denote the vacuum state by $|0\rangle=\prod_{j=0}^M|0\rangle_j$. 
As the action of the operator $B(u)$ (resp. $C(u)$) on a state increases 
(resp. decreases) 
the total number of particles
by one, we call the state $\prod_{j=1}^{N}B(u_{j})|0\rangle$  
the $N$-particle state and 
$\langle0|\prod_{j=1}^{N}C(u_{j})$ its conjugate. 
Obviously the $N$-particle state is a linear combination of $N$-particle bases. 
A matrix element $\langle \{n_{j}'\}|B(u)|\{n_{j}\}\rangle$ vanishes unless 
$|n_{j}'-n_{j}|\le 1$ and 
$N^{'}=N+1$. Similarly, a matrix element $\langle \{n_{j}'\}|C(u)|\{n_{j}\}\rangle$ 
vanishes unless  
$|n_{j}'-n_{j}|\le 1$ and $N^{'}=N-1$. Since the actions of $A(w)$ and $D(w)$ preserve 
the number of particles, 
matrix elements $\langle \{n_{j}'\}|A(w)|\{n_{j}\}\rangle, 
\langle \{n_{j}'\}|D(w)|\{n_{j}\}\rangle$ 
vanish unless  
$|n_{j}'-n_{j}|\le 1$ and $N^{'}=N$.

Now, we rephrase the arrow configuration of vertical graphs in the previous section 
as a non-intersecting up-right path configuration over the lattices. 
The action of the monodromy operator on a Fock state can be 
represented as a lattice path configuration on a vertical lattice. 

\begin{figure}[htbp]
  \begin{center}
{\setlength{\unitlength}{0.00083333in}
\begingroup\makeatletter\ifx\SetFigFont\undefined%
\gdef\SetFigFont#1#2#3#4#5{%
  \reset@font\fontsize{#1}{#2pt}%
  \fontfamily{#3}\fontseries{#4}\fontshape{#5}%
  \selectfont}%
\fi\endgroup%
{\renewcommand{\dashlinestretch}{30}
\scalebox{0.5}{
\begin{picture}(3644,6456)(0,-10)
\allinethickness{0.1cm}
\path(2422,6397)(2422,97)
\path(1222,6397)(1222,97)
\thinlines
\put(2422,3697){\blacken\ellipse{120}{120}}
\put(2422,3697){\ellipse{120}{120}}
\put(2422,1897){\blacken\ellipse{120}{120}}
\put(2422,1897){\ellipse{120}{120}}
\put(2422,2797){\blacken\ellipse{120}{120}}
\put(2422,2797){\ellipse{120}{120}}
\put(1222,4597){\blacken\ellipse{120}{120}}
\put(1222,4597){\ellipse{120}{120}}
\put(1222,3697){\blacken\ellipse{120}{120}}
\put(1222,3697){\ellipse{120}{120}}
\put(1222,1897){\blacken\ellipse{120}{120}}
\put(1222,1897){\ellipse{120}{120}}
\put(1222,2797){\blacken\ellipse{120}{120}}
\put(1222,2797){\ellipse{120}{120}}
\put(2422,5497){\blacken\ellipse{120}{120}}
\put(2422,5497){\ellipse{120}{120}}
\put(1222,5497){\blacken\ellipse{120}{120}}
\put(1222,5497){\ellipse{120}{120}}
\put(2422,997){\blacken\ellipse{120}{120}}
\put(2422,997){\ellipse{120}{120}}
\put(1222,997){\blacken\ellipse{120}{120}}
\put(1222,997){\ellipse{120}{120}}
\blacken\path(1222,1597)(1072,1372)(1372,1372)(1222,1597)
\path(1222,1597)(1072,1372)(1372,1372)(1222,1597)
\blacken\path(2422,5797)(2572,6022)(2272,6022)(2422,5797)
\path(2422,5797)(2572,6022)(2272,6022)(2422,5797)
\blacken\path(1222,4297)(1072,4072)(1372,4072)(1222,4297)
\path(1222,4297)(1072,4072)(1372,4072)(1222,4297)
\blacken\path(1222,3397)(1072,3172)(1372,3172)(1222,3397)
\path(1222,3397)(1072,3172)(1372,3172)(1222,3397)
\blacken\path(1222,4897)(1372,5122)(1072,5122)(1222,4897)
\path(1222,4897)(1372,5122)(1072,5122)(1222,4897)
\blacken\path(2422,5197)(2272,4972)(2572,4972)(2422,5197)
\path(2422,5197)(2272,4972)(2572,4972)(2422,5197)
\blacken\path(1222,2197)(1372,2422)(1072,2422)(1222,2197)
\path(1222,2197)(1372,2422)(1072,2422)(1222,2197)
\blacken\path(2422,2497)(2272,2272)(2572,2272)(2422,2497)
\path(2422,2497)(2272,2272)(2572,2272)(2422,2497)
\blacken\path(2422,3097)(2572,3322)(2272,3322)(2422,3097)
\path(2422,3097)(2572,3322)(2272,3322)(2422,3097)
\blacken\path(2422,1297)(2572,1522)(2272,1522)(2422,1297)
\path(2422,1297)(2572,1522)(2272,1522)(2422,1297)
\blacken\path(2422,3997)(2572,4222)(2272,4222)(2422,3997)
\path(2422,3997)(2572,4222)(2272,4222)(2422,3997)
\blacken\path(1222,397)(1372,622)(1072,622)(1222,397)
\path(1222,397)(1372,622)(1072,622)(1222,397)
\blacken\path(2422,697)(2272,472)(2572,472)(2422,697)
\path(2422,697)(2272,472)(2572,472)(2422,697)
\blacken\path(1222,6097)(1072,5872)(1372,5872)(1222,6097)
\path(1222,6097)(1072,5872)(1372,5872)(1222,6097)
\thinlines
\put(2422,4597){\blacken\ellipse{120}{120}}
\put(2422,4597){\ellipse{120}{120}}
\allinethickness{0.06cm}
{\color{red}{
\path(22,2947)(970,2947)(970,3622)(2200,3622)(3622,3622)
\path(22,997)(970,997)(970,2047)(2200,2047)(2200,2722)(3622,2722)
\path(22,3772)(970,3772)(970,4672)(2200,4672)(2200,5572)(3622,5572)
\path(22,2872)(3622,2872)
\path(22,5572)(970,5572)(970,6397)
\path(2200,22)(2200,847)(3622,847)
}}
\end{picture}
}}}
\end{center}
\caption{An example of a lattice path configuration 
			corresponding to an arrow configuration.}
  \label{fig3}
\end{figure}
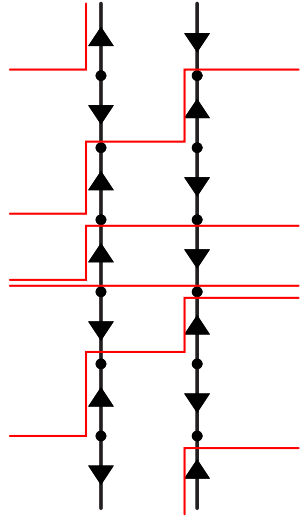

Let us fix an $N$-particle basis $\{n_{j}'\}$ and  an arrow configuration 
of a vertical graph. 
Put $n_{j}'$ horizontal paths entering the $j$-th site from the left of the lattice. 
Put also another path entering the $0$-th site from the bottom 
if the bottom arrow is inward. 
The rule is that only one path can go along an upward arrow and no path along 
a downward arrow. 
If the $j$-th site and the $(j+1)$-th site are connected by an upward arrow, one of the 
paths goes upward from the $j$-th site along the vertical lattice 
until it faces a downward arrow or another horizontal path, then it goes rightward. 
Other paths getting into the $j$-th site just pass rightward horizontally. 
A path entering the $M$-th site and going up along the top arrow, when the top arrow 
is upward, just goes vertically. 
Note that if there is no vertical path on a certain upward arrow, 
every $|\{n_{j}\}\rangle$ is annihilated by one of $\phi_j$'s.
As a result, we have a configuration $\{n_j\}$ by counting $n_j$ paths 
coming right out of the $j$-th site. 
When the monodromy operators act successively on a Fock state, 
we can construct up-right paths by repeating the above procedure. 
In this way, a lattice path configuration corresponds one-to-one 
to an arrow configuration of vertical graphs. 
Figure \ref{fig3} is an example of this correspondence.
% figure

We rewrite a Fock state in terms of a partition in order to obtain 
an explicit expression of the matrix elements of the monodromy operators. 
For a given $N$-particle basis $|\{n_{j}\}\rangle$ 
we assume that $n_{j_{r}}>0\ (r=1,\cdots, R)$ and $M\ge j_{1}>j_{2}>\cdots>j_{R}\ge 0$. 
The identification is that 
$\lambda=(\lambda_{1},\cdots,\lambda_{N})$, 
$M\ge\lambda_{1}\ge \lambda_{2}\cdots \ge \lambda_{N}\ge0$
where
\begin{eqnarray}
\begin{aligned}
&\lambda_{k}=j_{1},\qquad 1\le k\le n_{j_{1}} \\
&\lambda_{k}=j_{2},\qquad n_{j_{1}}+1\le k\le n_{j_{1}}+n_{j_{2}} \\
&\ \ \ \ \cdots  \\
&\lambda_{k}=j_{R},\qquad 
\sum_{r=1}^{R-1}n_{j_{r}}+1\le k\le \sum_{r=1}^{R}n_{j_{r}}=N.  
\end{aligned}
\end{eqnarray}
We write the $N$-particle basis $|\{n_{j}\}\rangle$ as $|\lambda\rangle$. 
The Fock space is spanned by the orthonormal basis $|\lambda\rangle$ of partitions. 
In a lattice path configuration of the monodromy operator, 
the site $\lambda_i$ is exactly the passing site of the $i$-th path from above. 

Then, the matrix elements of monodromy operators with respect to $|\lambda\rangle$ bases 
have the following expression in terms of the skew Schur functions $s_{\lambda/\mu}$:
\begin{proposition}
Matrix elements of the monodromy operators are given as follows $(\lambda\subseteq M^{N})$.
\begin{eqnarray}
\begin{aligned}
\langle\mu|A(u)|\lambda\rangle&=u^{-(M+1)}s_{\mu/\lambda}(u^{2}), \qquad
&&\mathrm{for}\ \mu\subseteq M^{N}, \\
\langle\mu|B(u)|\lambda\rangle&=u^{-M}s_{\mu/\lambda}(u^{2}), \qquad
&&\mathrm{for}\ \mu\subseteq M^{N+1}, \\
\langle\mu|C(u)|\lambda\rangle&=u^{M}s_{\lambda/\mu}(u^{-2}), \qquad
&&\mathrm{for}\ \mu\subseteq M^{N-1}, \\
\langle\mu|D(u)|\lambda\rangle&=u^{M+1}s_{\lambda/\mu}(u^{-2}), \qquad
&&\mathrm{for}\ \mu\subseteq M^{N}.
\end{aligned}
\label{formula-skew-operator}
\end{eqnarray}
\end{proposition}
\begin{proof}
We consider a non-vanishing $\langle\mu|C(u)|\lambda\rangle$. 
In the view of a lattice path, the partitions $\mu$ and $\lambda$ are 
identified with the initial and final positions of the up-right paths, respectively.
Fix $\mu\subseteq M^{N-1}$ and consider the corresponding path configuration 
with respect to $\lambda\subseteq M^N$. We treat $\lambda\subseteq M^N$ because
another path comes in along the bottom arrow for the configuration of $C(u)$. 

When, graphically, the arrow configuration for $C(u)$ is the basic one, 
the path configuration is such that 
all $N-1$ paths from the left pass through without going vertically, 
and another path comes in from the bottom and goes out at site $0$. 
Then, $\lambda$ is equal to $\mu$. 
Note that $\lambda_{N}=0$ is irrelevant to the shape $\lambda$. 
We find in this case $\langle\mu|C(u)|\lambda\rangle=u^{M}$. 

For the case where $\lambda\neq\mu$, 
the arrow configuration is the one where some downarrows except 
the boundary ones are flipped upside-down from the basic configuration. 
If one downarrow is flipped from the basic configuration, we have another factor 
$u^{-2}$. 
Let us consider a lattice path configuration 
where the $i$-th path (from above) goes along $l_{i}\ge0$ uparrows. 
There are $l=\sum l_{i}$ uparrows flipped upside-down from the basic configuration, and
we find $\langle\mu|C(u)|\lambda\rangle=u^{M-2l}$. 
Now, the partitions $\mu$ and $\lambda$ satisfy the following properties. 
First, $\lambda\supseteq\mu$ and $\lambda_{i}=\mu_{i}+l_{i}$. 
Second, the skew partition $\lambda/\mu$ is a one-vertical strip since 
the condition that only one path should pass through every uparrow tells 
$\lambda_{i+1}\le \mu_i\le \lambda_i,\ \forall i$.
From the definition 
of the skew Schur function, $s_{\lambda/\mu}(u^{-2})=u^{-2l}$. Together 
with these observations, we finally obtain 
$\langle\mu|C(u)|\lambda\rangle=u^{M}s_{\lambda/\mu}(u^{-2})$. 

Just similar considerations give matrix elements for the other operators. 
\end{proof}
By inserting the complete set $1=\sum_{\nu}|\nu\rangle\langle\nu|$ between 
every successive operators and using 
$\sum_{\nu} s_{\mu/\nu}(x)s_{\nu/\lambda}(y)=s_{\mu/\lambda}(x,y)$, 
we furthermore have the following relations. 
\begin{proposition}
The matrix elements of products of the monodromy operators are as follows 
($\lambda, \lambda'\subseteq M^{N}, \mu\subseteq M^{N+J}$).
\begin{eqnarray}
\begin{aligned}
\langle\mu|\prod_{j=1}^{J}B(u_{j})|\lambda\rangle&=&(u_{1}\cdots u_{J})^{-M}
s_{\mu/\lambda}(u_{1}^{2},\cdots,u_{J}^{2}),\\
\langle\lambda|\prod_{j=1}^{J}C(u_{j})|\mu\rangle&=&(u_{1}\cdots u_{J})^{M}
s_{\mu/\lambda}(u_{1}^{-2},\cdots,u_{J}^{-2}),\\
\langle\lambda^{'}|\prod_{j=1}^{J}A(u_{j})|\lambda\rangle&=&(u_{1}\cdots u_{J})^{-(M+1)}
s_{\lambda^{'}/\lambda}(u_{1}^{2},\cdots,u_{J}^{2}), \\
\langle\lambda|\prod_{j=1}^{J}D(u_{j})|\lambda^{'}\rangle&=&(u_{1}\cdots u_{J})^{M+1}
s_{\lambda^{'}/\lambda}(u_{1}^{-2},\cdots,u_{J}^{-2}).
\end{aligned}
\label{formula-skew-operator2}
\end{eqnarray}
Other elements are zero.
\end{proposition}
These expressions allow us to write a generalized scalar product as a sum of 
products of the skew Schur functions (see Section \ref{skewschur}).

The hermitian conjugates of the monodromy operators satisfy the following.  
\begin{proposition}The hermitian conjugates of the operators are related as
\begin{eqnarray}
\begin{aligned}
A^{\dagger}(u)&=D(u^{-1}), \qquad D^{\dagger}(u)=A(u^{-1}),\\
B^{\dagger}(u)&=C(u^{-1}), \qquad C^{\dagger}(u)=B(u^{-1}),
\end{aligned}
\label{hermite-operator}
\end{eqnarray}
i.e. for the monodromy matrix it holds that
\begin{eqnarray}
T^{\dagger}(u)=T(u^{-1}).
\end{eqnarray} 
\label{hermite-operator2}
\end{proposition}
\begin{proof}
From (\ref{formula-skew-operator}), we have 
$\langle\mu|A(u)|\lambda\rangle=u^{-(M+1)}s_{\mu/\lambda}(u^{2})=
\langle\lambda|D(u^{-1})|\mu\rangle$ 
and 
$\langle\mu|B(u)|\lambda\rangle=u^{-M}s_{\mu/\lambda}(u^{2})=
\langle\lambda|C(u^{-1})|\mu\rangle$. 
\end{proof}

\subsection{Generalized scalar product}\label{sec:GSP}
We briefly review the scalar product of the phase model. 
The state $\prod_{j=1}^{N} B(u_{j})|0\rangle$ is the 
$N$-particle state. 
The inner product of $N$-particle states is called a scalar product of 
the phase model. 
A scalar product of the phase model was first calculated in \cite{qboson2} 
through the partition function of the six-vertex model 
with domain wall boundary condition \cite{Kor82, Ize87}.  
Another way to calculate it is to use the Schur function representation of the 
$N$-particle state. The explicit expression of 
the scalar product of the $N$-particle state is~\cite{Bog}
\begin{eqnarray}
\langle0|\prod_{j=1}^{N}C(v_{j})\prod_{j=1}^{N}B(u_{j})|0\rangle
&=&\left(\prod_{i=1}^{N}\frac{u_{i}}{v_{i}}\right)^{M}\sum_{\lambda\subseteq M^{N}}
s_{\lambda}(u_{1}^{2},\cdots,u_{N}^{2})s_{\lambda}(v_{1}^{-2},\cdots,v_{N}^{-2}) 
\nonumber  \\
&=&\prod_{i<j}g(u_i,u_j) \prod_{i>j}g(v_i,v_j)
\det\big[H(v_{i},u_{j})\big]_{1\le i,j\le N},
\label{ScalarProduct}
\end{eqnarray}
where
\begin{eqnarray}
H(v_{i},u_{j})=\left\{\left(\frac{u_{j}}{v_{i}}\right)^{M+N}
-\left(\frac{v_{i}}{u_{j}}\right)^{M+N} \right\}
\times \frac{1}{\frac{u_{j}}{v_{i}}-\frac{v_{i}}{u_{j}}}.
\end{eqnarray}
To prove (\ref{ScalarProduct}), we just use the definition of the Schur function 
$s_{\lambda}(x)=\prod_{i>j}(x_{j}-x_{i})^{-1}\det x_{i}^{N-j+\lambda_{j}}$ 
when the length of $\lambda$ is equal to or less 
than $N$ and the Cauchy-Binet formula.

We define a generalized scalar product by the vacuum expectation value (VEV) of 
a sequence of the monodromy operators.  
It has a non-zero value only when the numbers of $B$ and $C$ are equal. 
In this article, we only consider the generalized scalar product of 
a $CB$-type word (see Section \ref{skew-word}):
\begin{align}
&\langle 0\vert
    \prod_{i=1}^{N_n} C(\l_i^{C_n})\prod_{i=1}^{M_n} A(\l_i^{A_n})\cdots
    \prod_{i=1}^{N_1} C(\l_i^{C_1})\prod_{i=1}^{M_1} A(\l_i^{A_1}) \nonumber\\
&\qquad\qquad\qquad
    \prod_{i=1}^{L_1} D(\l_i^{D_1})\prod_{i=1}^{K_1} B(\l_i^{B_1})\cdots
	\prod_{i=1}^{L_m} D(\l_i^{D_m})\prod_{i=1}^{K_m} B(\l_i^{B_m})
	\vert 0\rangle
\label{GSP}
\end{align}
A generalized scalar product of this type can be calculated in two ways: 
by use of the skew Schur function representation of 
the phase model, Proposition~\ref{formula-skew-operator2} (see Section \ref{skewschur}),
and by use of the commutation 
relations of the monodromy operators (see Section \ref{GSP-2}). 
In both cases, a generalized scalar product can be expressed in a determinant form. 

A scalar product of the phase model is related with the boxed plane partition 
by adjusting the parameters as shown in \cite{Bog}. 
Below, we show that a generalized scalar product is related with the 
boxed skew plane partition. 
A sequence of operators appeared in (\ref{GSP}) determines the base shape 
of the boxed skew plane partition. These subjects are considered in the successive 
sections.

\subsection{Inhomogeneous phase model}\label{IPM}
We can generalize the phase model to that with inhomogeneous parameters;  
the spectral parameter $u$ of $L_j$ is replaced by $uz_j$ in each lattice site $j$. 
Since $f(vz_j,uz_j)=f(v,u)$ and $g(vz_j,uz_j)=g(v,u)$, the $R$-matrix is
unchanged so that the relations for the monodromy matrix 
obtained from the QISM still remain the same. 
The only change that comes into our game is the functions,
\begin{align}
a(u;\{z_j\})=u^{-(M+1)}\prod_{j=0}^M z_j^{-1},\qquad
d(u;\{z_j\})=u^{M+1}\prod_{j=0}^M z_j.
\label{inhomo}
\end{align}
The calculation procedure for the generalized scalar product in this inhomogeneous case 
is much easier through the QISM than the explicit eigenfunctions 
in terms of the Schur function. Only we have to do is 
a replacement (\ref{inhomo}) in the final expressions obtained for the homogeneous case.

%%%%%%%% Plane Partition
\section{Plane Partition}
\label{sec:PP}

\subsection{Boxed plane partition}
A plane partition is an array of non-negative numbers 
\begin{eqnarray*}
\pi=(\pi_{ij}),\qquad i,j=1,2,\cdots,
\end{eqnarray*}
that is non-increasing in both $i$ and $j$. 
A plane partition is sometimes called a {\it 3D partition}. 
Piling $\pi_{ij}$ cubes over the $(i,j)$ square in the two-dimensional square 
lattice gives a three-dimensional object. 
The volume of $\pi$ is denoted by $|\pi|=\sum\pi_{ij}$. 
We call $\pi$ a boxed plane partition (or a plane partition in a box)
if $1\le i\le N, 1\le j\le L$ and $|\pi_{ij}|\le M$ for some $N,L$ and $M$. 
A boxed plane partition can be considered as 
stacks of cubes in a box with side lengths $N,L,M$. 
The generating function of the plane partition in 
an $N\times L\times M$-box is known to be 
\begin{eqnarray}
Z_{q}(N,L,M)=\sum_{\pi\in B(N,L,M)}q^{|\pi|}=
\prod_{i=1}^{N}\prod_{j=1}^{L}\frac{1-q^{i+j+M-1}}{1-q^{i+j-1}},
\label{genefuncpp}
\end{eqnarray}
where $B(N,L,M)$ stands for the $N\times L\times M$ box. 
If we take the limit $N,L,M\rightarrow\infty$ with the assumption $|q|<1$, 
the partition function $Z_{q}$ converges to the celebrated 
MacMahon function, $\prod_{n=1}^{\infty}(1-q^{n})^{-n}$.  

As explained in Section  \ref{LFYdiagram}, a scalar product of the phase model 
can be graphically interpreted in terms of lattice path configurations.
A lattice path configuration is also viewed as a $2D$-projection of 
a boxed (skew) plane partition through the lozenge tiling (see Section  \ref{LTSPP}); 
a plane partition in an $N\times L\times M$-box is also equivalent to 
a lozenge tiling of an $(N,L,M)$-semiregular hexagon. 
Hence there is a one-to-one correspondence between a boxed plane 
partition and a lattice path configuration. 

The generating function of the boxed skew plane partitions can be naturally obtained by 
a special choice of parameters of the generalized scalar product.  
Up to some factor, 
$\langle 0|\prod_{j=1}^{N}C(v_{j})\prod_{k=1}^{N}B(u_{k})|0\rangle$ 
with $u_j=q^{\frac{j-1}{2}},\ v_j=q^{-\frac{j}{2}},\ 1\le j\le N$ 
becomes the generating function of the plane partition in an $N\times N\times M$-box. 
Taking the corresponding generalized scalar product, 
an explicit determinant expression of (\ref{genefuncpp}) is
given in Section  \ref{genefuncbpp}.

\subsection{Boxed skew plane partition}
We introduce a skew shape $\lambda/\mu$ where $\mu\subseteq\lambda$. 
A skew plane partition $\pi$ whose shape is $\lambda/\mu$ is an array of non-negative 
integers $(\pi_{ij})$ defined on 
the $(i,j)$-squares contained in a skew shape $\lambda/\mu$ 
that is non-increasing in both $i$ and $j$. 
The volume of a skew plane partition is $|\pi|=\sum\pi_{ij}$. 
We call $\pi$ a boxed skew plane partition when 
$\lambda, \mu\subseteq N^{L}$ and $|\pi_{ij}|\le M$ for some $N, L$ and $M$.

%%%%%%%%%%%%% Skew PP & GSP
\section{Skew Plane Partition and Generalized Scalar Product}\label{SPP-GSP}

\subsection{Skew shape from a sequence of operators}\label{skew-word}
The scalar product of the $N$-particle states is relevant to the plane partition 
in an $N\times N\times M$-box. 
The base of the box is an $N\times N$ square, corresponding to a sequence of 
the operators $\prod_{j=1}^{N}C(v_{j})\prod_{j=1}^{N}B(u_{j})$. 
More generally, a sequence of the operators for the generalized scalar product reads  
the shape of the boxed skew plane partition, as we will explain below.

Let us introduce a {\it word} consisting of four letters $A,B,C$ and $D$. 
This is what ignores parameters of $A(u),\cdots,D(u)$ and vectors 
$\langle0|$ and $|0\rangle$ in the generalized scalar product. 
Let us define a $CB$-type word by a word in which the 
first $N$ letters are $A$ or $C$, and  
the next $N^{'}$ letters are $B$ or $D$ ($N$ and $N^{'}$ are non-negative integers). 
$CACADDBB$ is an example of a $CB$-type word. 

\begin{figure}[htbp]
  \begin{center}
{\setlength{\unitlength}{0.0008in}
\begingroup\makeatletter\ifx\SetFigFont\undefined%
\gdef\SetFigFont#1#2#3#4#5{%
  \reset@font\fontsize{#1}{#2pt}%
  \fontfamily{#3}\fontseries{#4}\fontshape{#5}%
  \selectfont}%
\fi\endgroup%
{\renewcommand{\dashlinestretch}{10}
\scalebox{0.5}{
\begin{picture}(9076,6000)(0,-10)
\put(1950,4200){\makebox(0,0)[lb]{{\SetFigFont{29}{38.4}{\rmdefault}{\mddefault}{\updefault}$\mu$}}}
\allinethickness{0.1cm}
\path(650,1665)(650,5400)(6130,5400)
{\color{red}{\path(4500,3750)(5100,3750)(5100,4500)(6150,4500)(6150,5400)(7900,5400)}
\path(600,600)(600,1650)(1800,1650)(1800,2400)(3150,2400)(3150,3300)(3750,3300)
\dashline{4}(3750,3300)(4500,3750)}
\path(7800,5410)(7810,600)(530,600)
\put(700,850){\makebox(0,0)[lb]{{\SetFigFont{29}{14.4}{\rmdefault}{\mddefault}{\updefault}$N_{n}$}}}
\put(800,1800){\makebox(0,0)[lb]{{\SetFigFont{29}{14.4}{\rmdefault}{\mddefault}{\updefault}$M_{n}$}}}
\put(1900,1800){\makebox(0,0)[lb]{{\SetFigFont{29}{14.4}{\rmdefault}{\mddefault}{\updefault}$N_{n-1}$}}}
\put(1900,2550){\makebox(0,0)[lb]{{\SetFigFont{29}{14.4}{\rmdefault}{\mddefault}{\updefault}$M_{n-1}$}}}
\put(3300,2700){\makebox(0,0)[lb]{{\SetFigFont{29}{14.4}{\rmdefault}{\mddefault}{\updefault}$N_{n-2}$}}}
\put(6600,5550){\makebox(0,0)[lb]{{\SetFigFont{29}{14.4}{\rmdefault}{\mddefault}{\updefault}$K_{m}$}}}
\put(4900,4650){\makebox(0,0)[lb]{{\SetFigFont{29}{14.4}{\rmdefault}{\mddefault}{\updefault}$K_{m-1}$}}}
\put(6300,4800){\makebox(0,0)[lb]{{\SetFigFont{29}{14.4}{\rmdefault}{\mddefault}{\updefault}$L_{m}$}}}
\put(5150,3800){\makebox(0,0)[lb]{{\SetFigFont{29}{14.4}{\rmdefault}{\mddefault}{\updefault}$L_{m-1}$}}}
\put(0,5650){\makebox(0,0)[lb]{{\SetFigFont{29}{14.4}{\rmdefault}{\mddefault}{\updefault}$(0,N+N_{D})$}}}
\put(7800,5700){\makebox(0,0)[lb]{{\SetFigFont{29}{14.4}{\rmdefault}{\mddefault}{\updefault}$(N+N_{A},N+N_{D})$}}}
\put(8000,200){\makebox(0,0)[lb]{{\SetFigFont{29}{14.4}{\rmdefault}{\mddefault}{\updefault}$(N+N_{A},0)$}}}
\put(0,0){\makebox(0,0)[lb]{{\SetFigFont{29}{14.4}{\rmdefault}{\mddefault}{\updefault}$(0,0)$}}}
\put(4950,1900){\makebox(0,0)[lb]{{\SetFigFont{29}{14.4}{\rmdefault}{\mddefault}{\updefault}$\lambda/\mu$}}}
\end{picture}
}}}
\end{center}
  \caption{The skew shape determined by a sequence of the monodromy operators.}
    \label{fig4}
\end{figure}
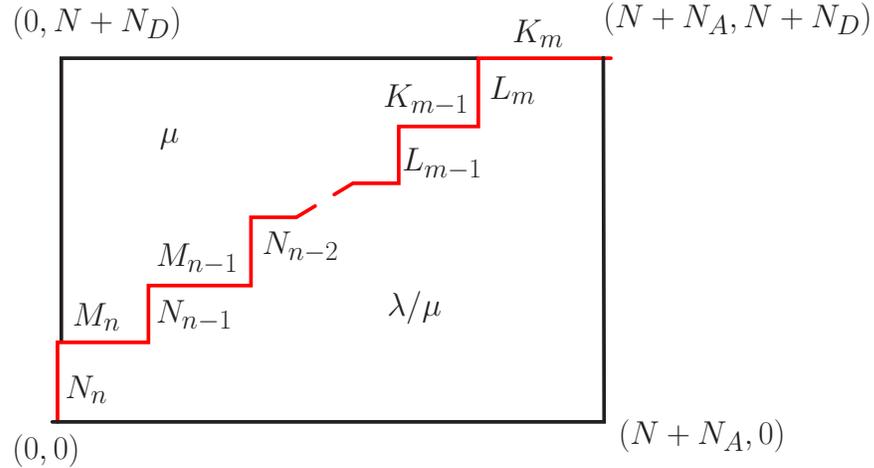

A generalized scalar product of a $CB$-type word corresponds to 
the plane partition with a skew shape $\lambda/\mu$ 
where $\lambda$ is a rectangle. 
We consider a $CB$-type word in which the numbers of $A,B,C$ and $D$ are $N_{A},N,N$ 
and $N_{D}$, respectively. Note that the numbers of $B$ and $C$ are equal to have 
a non-zero value for the generalized scalar product.   
The rectangle $\lambda$ turns out to be $(N+N_{D})\times(N+N_{A})$ (see Figure  \ref{fig4}). 
We put this rectangle on the two-dimensional lattice 
in a way that two diagonal corners are $(0,0)$ and $(N+N_{A},N+N_{D})$. 
Then, we draw a path as follows: 
1) a path starts from the origin, 
2) we read the word from left to right, and move one unit upward (resp. rightward) 
if the letter is either $C$ or $D$ (resp. either $A$ or $B$). 
The shape which is surrounded by two edges of the rectangle 
and the zig-zag path and contains the point $(0,N+N_{D})$ determines $\mu$. 
This is the one-to-one correspondence between a $CB$-type word and 
a boxed skew shape. Explicitly, for a generalized scalar product of a $CB$-type word 
(\ref{GSP}) the partition $\mu$ is given by 
\begin{eqnarray}
\begin{aligned}
\mu_{j}=&N+N_{A}-\sum_{p=1}^{s}K_{m+1-p}, 
&& 1+\sum_{p=1}^{s-1}L_{m+1-p}\le j\le\sum_{p=1}^{s}L_{m+1-p}, &&1\le s\le m, \\
\mu_{j}=&N+N_{A}-K-\sum_{q=1}^{r}M_{q}, 
&& L+1+\sum_{q=1}^{r-1}N_{q}\le j\le L+\sum_{q=1}^{r}N_{q}, &&1\le r\le n.
\label{basemu}
\end{aligned}
\end{eqnarray}
where we set $N=\sum_{i=1}^{n}N_{i},\ N_{A}=\sum_{i=1}^{n}M_{i}, \ 
N_{D}=\sum_{i=1}^{m}L_{i}$ 
and $K=\sum_{i=1}^{m}K_{i}=N$. 

% Figure

\subsection{Lozenge tiling and skew plane partition}\label{LTSPP}
In this subsection, we consider a map from an arrow configuration on vertical graphs 
in Section  \ref{LFYdiagram} to a lozenge tiling of a hexagon. 
A generalized scalar product with a special choice of the parameters can be 
naturally identified as the generating function of the skew plane partition. 

\begin{figure}[htbp]
  \begin{center}
{\setlength{\unitlength}{0.0005in} 
\begingroup\makeatletter\ifx\SetFigFont\undefined%
\gdef\SetFigFont#1#2#3#4#5{%
  \reset@font\fontsize{#1}{#2pt}%
  \fontfamily{#3}\fontseries{#4}\fontshape{#5}%
  \selectfont}%
\fi\endgroup%
{\renewcommand{\dashlinestretch}{30}
\scalebox{0.5}{
\begin{picture}(10995,4792)(0,-10)
\put(10000,0){\makebox(0,0)[lb]{{\SetFigFont{29}{38.4}{\rmdefault}{\mddefault}{\updefault}$c$}}}
\allinethickness{0.04cm}
\path(34,2306)(33,4706)
\path(44,4725)(2123,3525)
\path(2112,1106)(2112,3506)
\path(2112,1106)(34,2306)
\path(7569,2278)(5490,3477)
\path(3402,2295)(5479,3495)
\path(7569,2278)(5491,1076)
\path(5491,1076)(3413,2276)
\path(8884,3544)(8883,1144)
\path(10962,4744)(8884,3544)
\path(10962,4744)(10962,2344)
\path(8862,1125)(10941,2325)
\put(762,0){\makebox(0,0)[lb]{{\SetFigFont{29}{38.4}{\rmdefault}{\mddefault}{\updefault}$a$}}}
\put(5420,0){\makebox(0,0)[lb]{{\SetFigFont{29}{34.4}{\rmdefault}{\mddefault}{\updefault}$b$}}}
\allinethickness{0.08cm}
\path(1020,1710)(1020,4125)
\path(4383,1681)(6490,2881)     
\end{picture}
}}}
\end{center}
  \caption{Three types of tiles.}
  \label{fig5}
\end{figure}
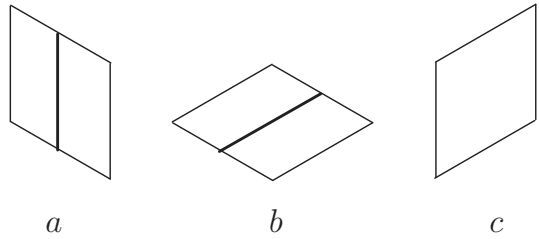

\begin{figure}[htbp]
  \begin{center}  
\begin{align*}
\pi=\left(
  \begin{array}{ccc}
       &  3  & 2   \\
      2 & 2   & 2   \\
     2 &  2  &  1  \\
  \end{array}
\right)
\hspace{8cm}
\end{align*}
{\setlength{\unitlength}{0.00083333in}
\begingroup\makeatletter\ifx\SetFigFont\undefined%
\gdef\SetFigFont#1#2#3#4#5{%
  \reset@font\fontsize{#1}{#2pt}%
  \fontfamily{#3}\fontseries{#4}\fontshape{#5}%
  \selectfont}%
\fi\endgroup%
{\renewcommand{\dashlinestretch}{30}
\scalebox{0.5}{
\begin{picture}
(6452,6134)(0,-10)
\put(6068,0){\makebox(0,0)[lb]{{\SetFigFont{34}{40.8}{\rmdefault}{\mddefault}{\updefault}B}}}
\put(68,0){\makebox(0,0)[lb]{{\SetFigFont{34}{40.8}{\rmdefault}{\mddefault}{\updefault}C}}}
\put(1193,0){\makebox(0,0)[lb]{{\SetFigFont{34}{40.8}{\rmdefault}{\mddefault}{\updefault}C}}}
\put(2468,0){\makebox(0,0)[lb]{{\SetFigFont{34}{40.8}{\rmdefault}{\mddefault}{\updefault}A}}}
\put(3668,0){\makebox(0,0)[lb]{{\SetFigFont{34}{40.8}{\rmdefault}{\mddefault}{\updefault}D}}}
\put(4868,0){\makebox(0,0)[lb]{{\SetFigFont{34}{40.8}{\rmdefault}{\mddefault}{\updefault}B}}}
\allinethickness{0.15cm}
\path(5018,6075)(5018,675)
\path(3818,6075)(3818,675)
\path(1418,6075)(1418,675)
\path(6218,6075)(6218,675)
\path(218,6075)(218,675)
\path(2618,6075)(2618,675)
\thinlines
\blacken\path(218,1275)(68,1050)(368,1050)(218,1275)
\path(218,1275)(68,1050)(368,1050)(218,1275)
\blacken\path(218,5475)(368,5700)(68,5700)(218,5475)
\path(218,5475)(368,5700)(68,5700)(218,5475)
\blacken\path(218,4575)(368,4800)(68,4800)(218,4575)
\path(218,4575)(368,4800)(68,4800)(218,4575)
\blacken\path(218,2175)(68,1950)(368,1950)(218,2175)
\path(218,2175)(68,1950)(368,1950)(218,2175)
\blacken\path(218,3075)(68,2850)(368,2850)(218,3075)
\path(218,3075)(68,2850)(368,2850)(218,3075)
\blacken\path(218,3750)(68,3975)(368,3975)(218,3750)
\path(218,3750)(68,3975)(368,3975)(218,3750)
\blacken\path(218,5475)(368,5700)(68,5700)(218,5475)
\path(218,5475)(368,5700)(68,5700)(218,5475)
\blacken\path(6218,975)(6368,1200)(6068,1200)(6218,975)
\path(6218,975)(6368,1200)(6068,1200)(6218,975)
\blacken\path(6218,5775)(6068,5550)(6368,5550)(6218,5775)
\path(6218,5775)(6068,5550)(6368,5550)(6218,5775)
\blacken\path(6218,2775)(6368,3000)(6068,3000)(6218,2775)
\path(6218,2775)(6368,3000)(6068,3000)(6218,2775)
\blacken\path(6218,1875)(6368,2100)(6068,2100)(6218,1875)
\path(6218,1875)(6368,2100)(6068,2100)(6218,1875)
\blacken\path(6218,3975)(6068,3750)(6368,3750)(6218,3975)
\path(6218,3975)(6068,3750)(6368,3750)(6218,3975)
\blacken\path(6218,4875)(6068,4650)(6368,4650)(6218,4875)
\path(6218,4875)(6068,4650)(6368,4650)(6218,4875)
\blacken\path(2618,1275)(2468,1050)(2768,1050)(2618,1275)
\path(2618,1275)(2468,1050)(2768,1050)(2618,1275)
\blacken\path(2618,5775)(2768,5550)(2468,5550)(2618,5775)
\path(2618,5775)(2768,5550)(2468,5550)(2618,5775)
\blacken\path(2618,3975)(2468,3750)(2768,3750)(2618,3975)
\path(2618,3975)(2468,3750)(2768,3750)(2618,3975)
\blacken\path(2618,2775)(2468,3000)(2768,3000)(2618,2775)
\path(2618,2775)(2468,3000)(2768,3000)(2618,2775)
\blacken\path(2618,4875)(2768,4650)(2468,4650)(2618,4875)
\path(2618,4875)(2768,4650)(2468,4650)(2618,4875)
\blacken\path(2618,2175)(2768,1950)(2468,1950)(2618,2175)
\path(2618,2175)(2768,1950)(2468,1950)(2618,2175)
\blacken\path(1418,5475)(1268,5700)(1568,5700)(1418,5475)
\path(1418,5475)(1268,5700)(1568,5700)(1418,5475)
\blacken\path(1418,2175)(1268,1950)(1568,1950)(1418,2175)
\path(1418,2175)(1268,1950)(1568,1950)(1418,2175)
\blacken\path(1418,3075)(1568,2850)(1268,2850)(1418,3075)
\path(1418,3075)(1568,2850)(1268,2850)(1418,3075)
\blacken\path(1418,3675)(1568,3900)(1268,3900)(1418,3675)
\path(1418,3675)(1568,3900)(1268,3900)(1418,3675)
\blacken\path(1418,4575)(1268,4800)(1568,4800)(1418,4575)
\path(1418,4575)(1268,4800)(1568,4800)(1418,4575)
\blacken\path(1418,1275)(1268,1050)(1568,1050)(1418,1275)
\path(1418,1275)(1268,1050)(1568,1050)(1418,1275)
\blacken\path(3818,5475)(3968,5700)(3668,5700)(3818,5475)
\path(3818,5475)(3968,5700)(3668,5700)(3818,5475)
\blacken\path(3818,4575)(3668,4800)(3968,4800)(3818,4575)
\path(3818,4575)(3668,4800)(3968,4800)(3818,4575)
\blacken\path(3818,1875)(3668,2100)(3968,2100)(3818,1875)
\path(3818,1875)(3668,2100)(3968,2100)(3818,1875)
\blacken\path(3818,3075)(3968,2850)(3668,2850)(3818,3075)
\path(3818,3075)(3968,2850)(3668,2850)(3818,3075)
\blacken\path(3818,975)(3968,1200)(3668,1200)(3818,975)
\path(3818,975)(3968,1200)(3668,1200)(3818,975)
\blacken\path(3818,3975)(3968,3750)(3668,3750)(3818,3975)
\path(3818,3975)(3968,3750)(3668,3750)(3818,3975)
\blacken\path(5018,2775)(4868,3000)(5168,3000)(5018,2775)
\path(5018,2775)(4868,3000)(5168,3000)(5018,2775)
\blacken\path(5018,4875)(5168,4650)(4868,4650)(5018,4875)
\path(5018,4875)(5168,4650)(4868,4650)(5018,4875)
\blacken\path(5018,975)(5168,1200)(4868,1200)(5018,975)
\path(5018,975)(5168,1200)(4868,1200)(5018,975)
\blacken\path(5018,1875)(5168,2100)(4868,2100)(5018,1875)
\path(5018,1875)(5168,2100)(4868,2100)(5018,1875)
\blacken\path(5018,3675)(5168,3900)(4868,3900)(5018,3675)
\path(5018,3675)(5168,3900)(4868,3900)(5018,3675)
\blacken\path(5018,5775)(4868,5550)(5168,5550)(5018,5775)
\path(5018,5775)(4868,5550)(5168,5550)(5018,5775)
\put(5018,4275){\blacken\ellipse{120}{120}}
\put(5018,4275){\ellipse{120}{120}}
\put(5018,3375){\blacken\ellipse{120}{120}}
\put(5018,3375){\ellipse{120}{120}}
\put(5018,1575){\blacken\ellipse{120}{120}}
\put(5018,1575){\ellipse{120}{120}}
\put(5018,2475){\blacken\ellipse{120}{120}}
\put(5018,2475){\ellipse{120}{120}}
\put(3818,4275){\blacken\ellipse{120}{120}}
\put(3818,4275){\ellipse{120}{120}}
\put(3818,3375){\blacken\ellipse{120}{120}}
\put(3818,3375){\ellipse{120}{120}}
\put(3818,1575){\blacken\ellipse{120}{120}}
\put(3818,1575){\ellipse{120}{120}}
\put(3818,2475){\blacken\ellipse{120}{120}}
\put(3818,2475){\ellipse{120}{120}}
\put(2618,4275){\blacken\ellipse{120}{120}}
\put(2618,4275){\ellipse{120}{120}}
\put(2618,3375){\blacken\ellipse{120}{120}}
\put(2618,3375){\ellipse{120}{120}}
\put(2618,1575){\blacken\ellipse{120}{120}}
\put(2618,1575){\ellipse{120}{120}}
\put(2618,2475){\blacken\ellipse{120}{120}}
\put(2618,2475){\ellipse{120}{120}}
\put(1418,4275){\blacken\ellipse{120}{120}}
\put(1418,4275){\ellipse{120}{120}}
\put(1418,3375){\blacken\ellipse{120}{120}}
\put(1418,3375){\ellipse{120}{120}}
\put(1418,1575){\blacken\ellipse{120}{120}}
\put(1418,1575){\ellipse{120}{120}}
\put(1418,2475){\blacken\ellipse{120}{120}}
\put(1418,2475){\ellipse{120}{120}}
\put(5003,5190){\blacken\ellipse{120}{120}}
\put(5003,5190){\ellipse{120}{120}}
\put(218,1575){\blacken\ellipse{120}{120}}
\put(218,1575){\ellipse{120}{120}}
\put(218,2475){\blacken\ellipse{120}{120}}
\put(218,2475){\ellipse{120}{120}}
\put(218,3375){\blacken\ellipse{120}{120}}
\put(218,3375){\ellipse{120}{120}}
\put(218,4275){\blacken\ellipse{120}{120}}
\put(218,4275){\ellipse{120}{120}}
\put(218,5175){\blacken\ellipse{120}{120}}
\put(218,5175){\ellipse{120}{120}}
\put(3818,5175){\blacken\ellipse{120}{120}}
\put(3818,5175){\ellipse{120}{120}}
\put(2618,5175){\blacken\ellipse{120}{120}}
\put(2618,5175){\ellipse{120}{120}}
\put(1418,5175){\blacken\ellipse{120}{120}}
\put(1418,5175){\ellipse{120}{120}}
\put(6218,1575){\blacken\ellipse{120}{120}}
\put(6218,1575){\ellipse{120}{120}}
\put(6218,2475){\blacken\ellipse{120}{120}}
\put(6218,2475){\ellipse{120}{120}}
\put(6218,3375){\blacken\ellipse{120}{120}}
\put(6218,3375){\ellipse{120}{120}}
\put(6218,4275){\blacken\ellipse{120}{120}}
\put(6218,4275){\ellipse{120}{120}}
\put(6218,5175){\blacken\ellipse{120}{120}}
\put(6218,5175){\ellipse{120}{120}}
{\color{red}{\allinethickness{0.15cm}
\path(8,600)(8,3475)(2393,3475)(2393,6075)
\path(1193,600)(1193,3325)(3668,3325)
	(3668,4350)(4793,4350)(4793,6075)
\path(2393,600)(2393,2375)(3668,2375)(3668,3225)
	(5993,3225)(5993,6075)}	}
\end{picture}
}}}\hspace{2cm}
   \scalebox{0.9}{\includegraphics{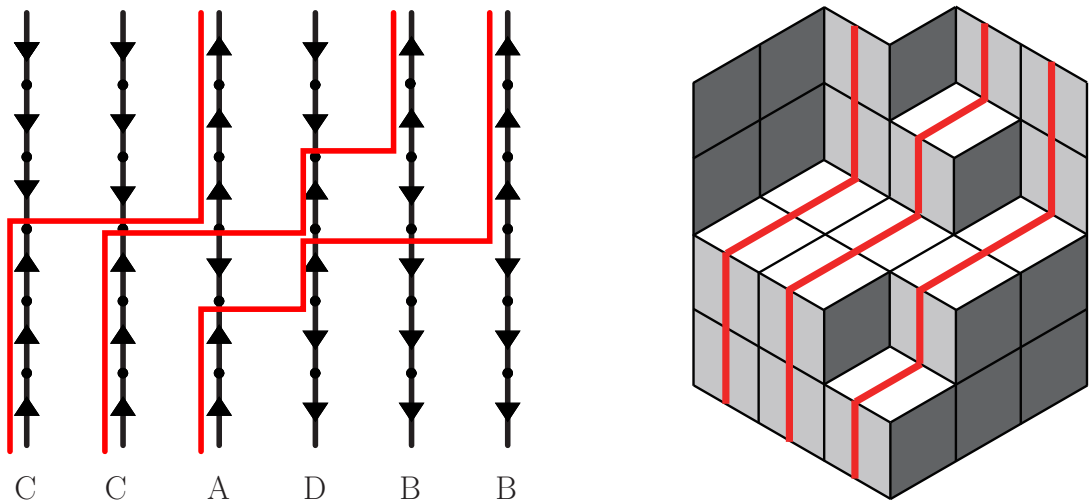}}
  \end{center}
  \caption{An example of identification of a skew plane partition with 
			a lattice path configuration and a lozenge tiling.}
  \label{fig6}
\end{figure}

Let us consider a lattice path configuration of a $CB$-type word. 
We arrange a lozenge tiling from this lattice path configuration. 
The domain of tiles is determined by the shape of the corresponding skew plane partition.
Let one unit of a path going upward correspond to a tile of type $a$, 
and let one unit of a path going rightward correspond to 
a tile of type $b$ in Figure \ref{fig5}.  
The remained spaces are filled with tiles of type $c$. 
% Figure
Figure \ref{fig6} is an example of the correspondence among a lattice path configuration, 
an arrow configuration and a lozenge tiling. 
We also draw paths on the lozenge tiling to help understanding. 
These paths on a lozenge tiling are indeed non-intersecting by construction. 
We can see that only non-intersecting lattice path configurations are mapped 
to lozenge tilings. 

A lozenge tiling gives a $2D$-projection of a $3D$ piling of cubes in a box, 
therefore, a boxed skew plane partition. 
% Figure

In the arrow graph representation, we can show that the generalized scalar product 
of a $CB$-type word (\ref{GSP}) is written as
\begin{eqnarray}
\sum 
\prod_c\prod^{N_{c}}_{n=1}(u^{C_{c}}_{n})^{M-2t^{n}_{\uparrow}}
\prod_a\prod^{M_{a}}_{m=1}(u^{A_{a}}_{m})^{(M-1)-2t^{m}_{\uparrow}}
\prod_b\prod^{L_{b}}_{l=1}(u^{B_{b}}_{l})^{2t^{l}_{\downarrow}-M}
\prod_d\prod^{K_{d}}_{k=1}(u^{D_{d}}_{k})^{2t^{k}_{\downarrow}-(M-1)}
\label{sum-GSP}
\end{eqnarray}
where the summation is taken over all arrow configurations, 
and $t^{i}_{\uparrow}$ (resp. $t^{i}_{\downarrow}$) counts
the number of up (resp. down) arrows except the boundary arrows 
in the $i$-th vertical graph. 
The powers of parameters are obtained by flipping some arrows upside-down from the 
basic configurations. 
Since the arrows at the boundaries are different, the powers of $u^{C}$ and $u^{A}$ 
(or $u^{D}$ and $u^{B}$) are different by one. 

Suppose that $\lambda$ is an $(N+N_{D})\times (N+N_{A})$ rectangle, 
$\mu$ is a partition (\ref{basemu}) and 
$\pi$ is a skew plane partition whose shape is $\lambda/\mu$.  
Using the correspondence of a skew plane partition and an arrow configuration, we obtain 
\begin{eqnarray}
|\pi|+M|\mu|+\frac{M}{2}(N_{A}-N_{D})(N_{A}-N_{D}+1)
=\sum_{j=1}^{N+N_{A}}(-j)(-t_{\uparrow}^{j})
+\sum_{j=1}^{N+N_{D}}(j-1)t_{\downarrow}^{j}.
\label{spp-gsp1}
\end{eqnarray}

For $1\le i\le n$, let $s(A_{i})$ be the set of $M_{i}$ successive integers from 
$\sum_{j=1}^{i-1}M_{j}+\sum_{j=1}^{i-1}N_{j}+1$, $s(A)=\cup_{i=1}^{n}s(A_{i})$,
and let $s(C_{i})$ be the set of $N_{i}$ successive integers from 
$\sum_{j=1}^{i}M_{j}+\sum_{j=1}^{i-1}N_{j}+1$, $s(C)=\cup_{i=1}^{n}s(C_{i})$. 
Note that $s(A)\cup s(C)=\{1,2,\cdots, N+N_{A}\}$.
This decomposition of integers corresponds to the product chain
$\prod^{N_{n}}C\prod^{M_{n}}A\cdots\prod^{N_{1}}C\prod^{M_{1}}A$. 
Similarly, for $1\le i\le m$, let $s(D_{i})$ be the set of $L_{i}$ successive integers 
from 
$\sum_{j=1}^{i-1}L_{j}+\sum_{j=1}^{i-1}K_{j}+1$, $s(D)=\cup_{i=1}^{m}s(D_{i})$,
and let $s(B_i)$ be the set of $K_i$ successive integers from 
$\sum_{j=1}^{i}L_{j}+\sum_{j=1}^{i-1}K_{j}+1$, $s(B)=\cup_{i=1}^{m}s(B_{i})$. 
Note the relations,
\begin{eqnarray}
\begin{aligned}
\sum_{j\in s(C)}j+\sum_{j\in s(B)}(j-1)&=|\lambda/\mu|, \\
\sum_{j\in s(A)}j+\sum_{j\in s(D)}(j-1)&=|\mu|+\frac{1}{2}(N_{A}-N_{D})(N_{A}-N_{D}+1).
\end{aligned}\label{spp-gsp2}
\end{eqnarray}
Then from (\ref{spp-gsp1}), we can show the following relation. 
\begin{eqnarray}
&&\sum_{j\in s(C)}(-\frac{j}{2})(M-2t_{\uparrow}^{j})+
\sum_{j\in s(A)}(-\frac{j}{2})(M-1-2t_{\uparrow}^{j}) \nonumber\\
&&\qquad\qquad
+\sum_{j\in s(B)}(\frac{j-1}{2})(2t_{\downarrow}^{j}-M)
+\sum_{j\in s(D)}(\frac{j-1}{2})(2t_{\downarrow}^{j}-M+1) \nonumber\\[0.3cm]
&&=|\pi|+\frac{2M+1}{2}|\mu|-\frac{M}{2}|\lambda|+
\frac{M+1}{4}(N_{A}-N_{D})(N_{A}-N_{D}+1).
\label{formula-spp-gsp}
\end{eqnarray}
We parameterize $u^{A},u^{B},u^{C}$ and $u^{D}$ as
\begin{eqnarray}
\begin{aligned}
\{u^{A_{i}}\}=\{q^{-\frac{j}{2}}|j\in s(A_{i})\}, \qquad &\mbox{for}\ 1\le i\le n\\
\{u^{B_{i}}\}=\{q^{\frac{j-1}{2}}|j\in s(B_{i})\}, \qquad &\mbox{for}\ 1\le i\le m\\
\{u^{C_{i}}\}=\{q^{-\frac{j}{2}}|j\in s(C_{i})\}, \qquad &\mbox{for}\ 1\le i\le n\\
\{u^{D_{i}}\}=\{q^{\frac{j-1}{2}}|j\in s(D_{i})\}, \qquad &\mbox{for}\ 1\le i\le m.
\end{aligned}
\label{parameter-q}
\end{eqnarray}
We denote by $Z_{GSP}(q^{\bullet})$ the generalized scalar product (\ref{GSP}) 
with the parameterization (\ref{parameter-q}). 
Combining (\ref{sum-GSP}) and (\ref{formula-spp-gsp}), 
we find the generating function of a boxed skew plane partition is equal to 
\begin{eqnarray*}
q^{-\alpha} Z_{GSP}(q^{\bullet}),
\label{spp-gsp-relation}
\end{eqnarray*}
where 
\begin{eqnarray*}
\alpha=
\frac{2M+1}{2}|\mu|-\frac{M}{2}|\lambda|+\frac{M+1}{4}(N_{A}-N_{D})(N_{A}-N_{D}+1).
\end{eqnarray*}

We give a remark about the upside-down property.  
The hermitian conjugate of a generalized scalar product of a $CB$-type word 
yields the generating function of another kind of plane partition
whose base shape is a hook Young diagram. 
From Proposition \ref{hermite-operator2}, the hermitian conjugate of the monodromy operators 
make the arrow graphs upside-down and make a lattice path configuration and 
a lozenge tiling rotate 180 degrees.
See Figure  \ref{fig6} rotated 180 degrees.

%%%%%%%%%%%% GSP via skew Schur
\section{Genelarized Scalar Product via Skew Schur functions }\label{skewschur}

\subsection{Cauchy-Binet formula}\label{Cauchy-Binet}
In the subsequent, we present a formula for the generalized scalar product 
that is a determinant of the skew Schur functions. 
For this purpose, the basic is the Cauchy-Binet formula for calculating determinant. 

Let two matrices $A$ and $B$ be of size $m\times n$ and $n\times m$, respectively 
(without loss of generality, $n\ge m$).
Let $\mathcal{S}_{mn}$ denote the set of strictly increasing sequences of length $m$ 
that can be chosen from $\{1,2,\cdots,n\}$.
For any matrix $X$ of size $n\times m$ and any $S=\{s_{1},\cdots,s_{m}\}\in \calS_{mn}$, 
we denote by $X(S|m)$ the $m\times m$
matrix obtained from $X$ by choosing all $m$ columns of $X$ and 
the $m$ rows numbered by $S$, $i.e.$ $(X(S|m))_{ij}=X_{s_{i}j}$. 
Similarly, if $X$ is $m\times n$, denote the $m\times m$ matrix $X_{is_{j}} 
(1\le i,j\le m)$ by $X(m|S)$. The Cauchy-Binet formula shows that
the determinant of the product of two rectangular matrices $A$ and $B$ is 
\begin{eqnarray}
\det(AB)=\sum_{S\in\mathcal{S}_{mn}}\det(B(S|m))\det(A(m|S)).
\label{CBformula}
\end{eqnarray}

\subsection{Skew Schur functions}
For any partition $\lambda=(\lambda_{1},\cdots,\lambda_{N}), \lambda_{i}\ge 0$
and some non-negative integer $L$, we define 
$\tilde{\lambda}=(\tilde{\lambda}_{1},\cdots,\tilde{\lambda}_{N+L})$ as
\begin{eqnarray*}
\tilde{\lambda}_{i}=\left\{
  \begin{array}{lc}
  \tilde{\lambda}_{i}=i,   &   1\le i\le L    \\
    \tilde{\lambda}_{i}=\lambda_{N+L+1-i}+i,    &   L+1\le i\le N+L 
  \end{array}
\right. .
\end{eqnarray*}
For any partitions $\lambda=(\lambda_{1},\cdots,\lambda_{N})$ and $\mu\subseteq\lambda$, 
the skew Schur function has a 
determinant expression of the form ({\it Jacobi-Trudi identity}) 
\begin{eqnarray}
s_{\lambda/\mu}&=&\det\left(h_{\lambda_{i}-\mu_{j}-i+j}\right)_{1\le i,j\le N} 
\nonumber\\
&=&\det\left(h_{\tilde{\lambda}_{i}-\tilde{\mu}_{j}}\right)_{1\le i,j\le N+L}.
\end{eqnarray}
where $h_n,\ n\in\mathbb{Z}$ is the complete symmetric function.
Applying the Cauchy-Binet formula (\ref{CBformula}), we can show the following expressions. 
\begin{proposition}\label{formula-ss}
For non-negative integers $N$, $L$ and $M$, we have 
\begin{eqnarray}
\sum_{\lambda\subseteq M^{N}}s_{\lambda/\mu}(x)s_{\lambda/\nu}(y)
&=&\det\left[\sum_{l=1}^{N+M+L}h_{l-\tilde{\mu}_{i}}(x)
h_{l-\tilde{\nu}_{j}}(y)\right]_{1\le i,j\le N+L},\\
\sum_{\lambda\subseteq M^{N}}s_{\mu/\lambda}(x)s_{\nu/\lambda}(y)
&=&\det\left[\sum_{l=1}^{N+M+L}h_{-l+\tilde{\mu}_{i}}(x)
h_{-l+\tilde{\nu}_{j}}(y)\right]_{1\le i,j\le N+L}.
\end{eqnarray}
More generally, 
let $\lambda^{i}=(\lambda^{i}_{1},\cdots,\lambda^{i}_{N_{i}}),\ 0\le i\le 2n$ 
and let $x^j=(x^{j}_{1},\cdots,x^{j}_{L_{j}}),\ 1\le j\le 2n$ denote $2n$ sets of 
variables, then we have 
a summation of products of the skew Schur functions over 
$\sigma=\{ \lambda^{i}| \lambda^{i}_{1}\le M,\ 1\le i\le 2n-1\}$ as
 a determinant of complete symmetric functions. 
\begin{eqnarray}
\sum_{\sigma}\prod_{i=0}^{n-1}
s_{\lambda^{2i+1}/\lambda^{2i}}(x^{2i+1})
s_{\lambda^{2i+1}/\lambda^{2i+2}}(x^{2i+2}) 
=\det\left[ 
A_{\tilde{\lambda}^{0}_{i}\tilde{\lambda}^{2n}_{j}}(x^{1},\cdots,x^{2n})
\right]_{1\le i,j\le N+L},  
\end{eqnarray}
where $N=\max\{ N_{i}\vert 1\le i\le 2n\}, L=\sum L_{i}$, and 
\begin{eqnarray}
A_{l_{0}l_{2n}}(x^{1},\cdots,x^{2n})&=&\sum_{l_{1},\cdots,l_{2n-1}=1}^{N+L+M}
\prod_{k=0}^{n-1}h_{l_{2k+1}-l_{2k}}(x^{2k+1})h_{l_{2k+1}-l_{2k+2}}(x^{2k+2}). 
\end{eqnarray}
\end{proposition}

Let us take the limit $M\to\infty$.  
The generating function of $A_{ij}(x)$ becomes
\begin{eqnarray}
\phi^{A}(x;z,w)&\equiv&
\sum_{i,j=1}^{\infty}z^{i}w^{j}A_{ij}(x^{1},\cdots,x^{2n}) \nonumber\\
&=&\bigg[\frac{wz}{1-wz}\prod_{i=1}^{n}\prod_{j}(1-x^{i}_{j}z^{-\sigma(i)})^{-1}
\prod_{i=n+1}^{2n}\prod_{j}(1-x^{i}_{j}w^{\sigma(i)})^{-1}\bigg]_{+}
\end{eqnarray}
where we take only positive powers and $\sigma(i)=(-1)^{i+1}$. 
Here, suppose that two functions $(1-\xi \zeta^{\pm 1})^{-1},\ \zeta=z,w$ represent 
the formal power series of $\zeta$: 
$(1-\xi \zeta^{\pm 1})^{-1}=\sum_{j=0}^{\infty}\xi^{j}\zeta^{\pm j}$.
If we also take partitions $\lambda^{0}, \lambda^{2n}=\emptyset$, we have
\begin{eqnarray}
\sum_{\lambda^{1},\cdots,\lambda^{2n-1}}
\prod_{i=0}^{n-1}s_{\lambda^{2i+1}/\lambda^{2i}}(x^{2i+1})
s_{\lambda^{2i+1}/\lambda^{2i+2}}(x^{2i+2})
&=&\det \big[A_{ij}(x^{1},\cdots,x^{2n-1})\big]_{1\le i,j\le N+L} \nonumber\\
&=&\prod_{i>j}\prod_{a>b}(1-x^{i}_{a}x^{j}_{b})^{-1}.
\label{schursum}
\end{eqnarray}

\subsection{Skew Schur function representation of generalized scalar products}
Recall that matrix elements of the monodromy operators are expressed 
in terms of the skew Schur functions (\ref{formula-skew-operator}) 
and (\ref{formula-skew-operator2}). 
From Proposition \ref{formula-ss}, 
the generalized scalar product of the form (\ref{GSP})
is written as
\begin{multline*}
\prod_{j=1}^n
\frac{(\prod u^{C_{j}})^{M}}{(\prod u^{A_{j}})^{M+1}}
\prod_{k=1}^m
\frac{(\prod u^{D_{k}})^{M+1}}{(\prod u^{B_{k}})^{M}}
\sum_{\lambda_1^i\le M,\ 1\le i\le 2(n+m)-1}
s_{\lambda^{1}}((u^{C_{n}})^{-2})s_{\lambda^{1}/\lambda^{2}}((u^{A_{n}})^{2})
\cdots  \\
\times s_{\lambda^{2n-1}/\lambda^{2n-2}}((u^{C_{1}})^{-2})s_{\lambda^{2n-1}/\lambda^{2n}}((u^{A_{1}})^{2})
\times s_{\lambda^{2n+1}/\lambda^{2n}}((u^{D_{1}})^{-2})s_{\lambda^{2n+1}/\lambda^{2n+2}}((u^{B_{1}})^{2}) 
\cdots\\
\times s_{\lambda^{2(n+m)-1}/\lambda^{2(n+m)-2}}((u^{D_{m}})^{-2})s_{\lambda^{2(n+m)-1}}((u^{B_{m}})^{2}) \\[0.5cm]
=
\prod_{j=1}^n
\frac{(\prod u^{C_{j}})^{M}}{(\prod u^{A_{j}})^{M+1}}
\prod_{k=1}^m
\frac{(\prod u^{D_{k}})^{M+1}}{(\prod u^{B_{k}})^{M}}
\det \bigg[
A_{ij}\left(
(u^{C_{n}})^{-2},(u^{A_{n}})^{2},\cdots,(u^{D_{m}})^{-2},(u^{B_{m}})^{2}\right)
\bigg]_{1\le i,j\le N+L}
\end{multline*}
where 
$(u^{C_{n}})^{-2}=((u_{1}^{C_{n}})^{-2},\cdots,(u_{N_{n}}^{C_{n}})^{-2})$ and so on.
In particular, consider the plane partition in an $N\times L\times \infty$ -box.
This corresponds to $\langle 0\vert \prod_{k=L-N+1}^L C(q^{-k/2})$ $\prod_{k=1}^{L-N} A(q^{-k/2})$
$\prod_{k=1}^N B(q^{(k-1)/2}) \vert 0\rangle$. 
If we substitute the parameterization (\ref{parameter-q}) into the above relation, 
we may reproduce (\ref{genefuncpp}) 
with $M\to\infty$ by the help of (\ref{schursum}), as expected.

%%%%%%%%%%%%%
% Section for the direct proof 
%%%%%%%%%%%%5
\section{Direct Calculation of Generalized Scalar Products}  \label{GSP-2}
%%% $\langle 0\vert\prod C\prod A\cdots\prod D\prod B\cdots\vert 0\rangle$

In the previous section, we have established a determinant formula for the 
generalized scalar product in terms of the skew Schur functions. 
In this section, we give a more explicit determinant formula. 
This is done by a direct application of the relations obtained from 
the QISM in Section  \ref{phase-YBR}. 
The advantage of the formula here is a simplicity for evaluations. 
Actually in the next section, we demonstrate an evaluation of the determinant 
corresponding to the boxed plane partition in an $N\times L\times M$-box. 

We remark about the notations. 
In this section, we often drop the indices of variables  
to avoid notational complexity if there is no confusion. 
For instance, we abbreviate 
$\prod_{i=1}^N K(\l_i^A)$ to $\prod^N K(\l^A)$ or just $\prod K(\l^A)$ 
and so on. 
Define the functions 
$\D(u,v)=u^2-v^2$, $a(u)=u^{-(M+1)}$ and $d(u)=u^{M+1}$. 

\begin{theorem}
Generalized scalar products have the following expression in terms of 
a determinant. 
\begin{align}
&\langle 0\vert
    \prod_{i=1}^{N_n} C(\l_i^{C_n})\prod_{i=1}^{M_n} A(\l_i^{A_n})\cdots
    \prod_{i=1}^{N_1} C(\l_i^{C_1})\prod_{i=1}^{M_1} A(\l_i^{A_1}) \nonumber\\
&\qquad\qquad\qquad
    \prod_{i=1}^{L_1} D(\l_i^{D_1})\prod_{i=1}^{K_1} B(\l_i^{B_1})\cdots
	\prod_{i=1}^{L_m} D(\l_i^{D_m})\prod_{i=1}^{K_m} B(\l_i^{B_m})
	\vert 0\rangle\nonumber\\
&=\prod_{i<j}g(\l_i^{B\vee D},\l_j^{B\vee D}) 
\prod_{i>j}g(\l_i^{A\vee C},\l_j^{A\vee C})
I_n\left(\{\l^C\};\{\l^A\}\right) 
J_m\left(\{\l^D\};\{\l^B\}\right)\nonumber\\
&\qquad
(-1)^{LM}
\mathrm{det}\ G\left(\{\l^{A\vee C}\};\{\l^{B\vee D}\}\right)
\label{detformula}
\end{align}
where $G=\left(G_{i,j}^{a,b}\right)$ is an $(N+M+L)\times(N+M+L)$ matrix with
\begin{align}
G_{i,j}^{a,b}=
\left\{
\begin{array}{l}
H(\l_i^{A\vee C},\l_j^{B\vee D}), \qquad\qquad\qquad 
\begin{array}[t]{l}
a=0,\ b=0,\\
1\le i\le N+M,\ 1\le j\le K+L(=N+L)
\end{array}
\\
\\
h_{b,j}(\l_i^{A\vee C}), \qquad\qquad\qquad 
\begin{array}[t]{l}
a=0,\ 1\le b\le n,\\
1\le i\le N+M,\ 1\le j\le M_b
\end{array}
\\
\\
\tilde{h}_{a,i}(\l_j^{B\vee D}), \qquad\qquad\qquad 
\begin{array}[t]{l}
1\le a\le m,\ b=0,\\
1\le i\le L_a,\ 1\le j\le K+L
\end{array}
\\
\\
0, \qquad\qquad\qquad \qquad\qquad
\begin{array}[t]{l}
1\le a\le m,\ 1\le b\le n, \\
1\le i\le L_a,\ 1\le j\le M_b
\end{array}
\end{array}
\right.
\end{align}
and the functions are
\begin{eqnarray}
\begin{aligned}
&h_{b,j}(\l)=\prod_{\ell=1}^{b-1}\left[
\prod\D(\l^{C_\ell},\l)\D(\l^{A_\ell},\l)\right]
a(\l) \l^{-2(j-1)-2\sum_{\ell=1}^{b-1}M_\ell-N}
\\
&
\tilde{h}_{a,i}(\l)=
\prod_{\ell=1}^{a-1} \left[
\prod \D(\l,\l^{B_\ell})\D(\l,\l^{D_\ell})\right]
d(\l) \l^{2(i-1)+\sum_{\ell=a}^mK_\ell-\sum_{\ell=1}^{a-1}K_\ell}
\end{aligned}
\end{eqnarray}
and
\begin{eqnarray}
\begin{aligned}
&I_n\left(\{\l^{C_1}\},\cdots,\{\l^{C_n}\};
\{\l^{A_1}\},\cdots,\{\l^{A_n}\}\right)\\
&\qquad
=\prod_{i=1}^n \left[ 
		\left(\prod\l^{C_i}\right)^{\sum_{j=1}^i M_j-\sum_{j=i+1}^n M_j}
		\left(\prod\l^{A_i}\right)^{\sum_{j=1}^i M_j-\sum_{j=i+1}^n M_j-1}
	\right]
\\
&J_m\left(\{\l^{D_1}\},\cdots,\{\l^{D_m}\};
\{\l^{B_1}\},\cdots,\{\l^{B_m}\}\right) \\
&\qquad
=\prod_{i=1}^m \left[
	\left(\prod \l^{D_i}\right) ^{-\sum_{j=1}^n L_j+1}
	\left(\prod \l^{B_i}\right) ^{-\sum_{j=1}^n L_j}
	\right] .
\end{aligned}
\end{eqnarray}
Here we set
$N=\sum N_i$, $M=\sum M_i$, $K=\sum K_i =N$ and $L=\sum L_i$. 
The sequence $\{\l_i^{A\vee C}\}_{i=1}^{N+M}$ is arranged as 
$\{\l_i^{C_n}\}_{i=1}^{N_n}$ followed by 
$\{\l_i^{A_n}\}_{i=1}^{M_n},$ $\cdots,$ $\{\l_i^{C_1}\}_{i=1}^{N_1},$ 
$\{\l_i^{A_1}\}_{i=1}^{M_1}$. 
Similarly, the sequence $\{\l_i^{B\vee C}\}_{i=1}^{K+L}$ is arranged as
$\{\l_i^{B_m}\}_{i=1}^{K_m}$ followed by 
$\{\l_i^{D_m}\}_{i=1}^{L_m},$ $\cdots,$ $\{\l_i^{B_1}\}_{i=1}^{K_1}, $
$\{\l_i^{D_1}\}_{i=1}^{L_1}$. 
\end{theorem}

\begin{proof}
Firstly, since (\ref{AA}) holds, the generalized scalar product is symmetric 
in each of $\{\l^{A_i}\}$, $\{\l^{B_i}\}$, $\{\l^{C_i}\}$ and $\{\l^{D_i}\}$,
respectively. 
We can generalize the relations (\ref{CA1}) and (\ref{CA2}) to
\begin{align}
&\prod^N C(\l^C) \prod^M A(\l^A)\nonumber\\
&=
\sum_{A',C'}\left(\prod\D(\l^{C'},\l^{A'})\right) ^{-1}
\left(\prod\l^{C'}\right)^{2M-1}
\left(\prod\l^C\right)
\prod^M A(\l^{A'}) \prod^N C(\l^{C'}).
\label{CCAA}
\end{align}
Here the summation is taken over all decompositions 
$\{\l_i^{C'}\}_{i=1}^N \cup \{\l_i^{A'}\}_{i=1}^M 
=\{\l_i^{C}\}_{i=1}^N \cup \{\l_i^{A}\}_{i=1}^M$. 
The relation (\ref{CCAA}) is obtained as follows. 
First, the coefficient of $\prod A(\l^A) \prod C(\l^C)$ is clearly
$\prod f(\l^A,\l^C)$. 
To see the coefficient of $\prod A(\l^{A'}) \prod C(\l^{C'})$, 
exchange the variables of $\prod C(\l^C)$ $\prod A(\l^A)$ 
by relation (\ref{CA1}) till $\prod C(\l^{C'}) \prod A(\l^{A'})$, 
then move the $A$'s through $C$'s and we have a coefficient 
\[\prod_i \frac{g(\l_i^{CA'},\l_i^{AC'})}{f(\l_i^{CA'},\l_i^{AC'})}
\prod f(\l^{A'},\l^{C'})\] 
where each $\l_i^{CA'}\in\{\l^C\}\cap\{\l^{A'}\}$ and $\l_i^{AC'}\in\{\l^A\}\cap\{\l^{C'}\}$ appears once in the product. 
By the definition (\ref{fg}) of $f$ and $g$, we arrive at (\ref{CCAA}). 

Fix the number of the operators $C$ and $B$ both $N$. 
For the case where there is no $A$'s and no $D$'s, the formula is nothing but 
the scalar product (\ref{ScalarProduct}). 
Assume the formula for $n$ and $m$, that is, $n$ products of $A$'s 
and $m$ products of $D$'s are inserted. 
Then we calculate the case of $n+1$ insertion of products of $A$'s. 
Here we set $N'=N_1+\cdots+N_n$, $M'=M_1+\cdots+M_n$. 
From (\ref{CCAA}) and the formula for $n$ and $m$, we get
\begin{align}
&\langle 0\vert
    \prod^{N_{n+1}} C(\l^{C_{n+1}})\prod^{M_{n+1}} A(\l^{A_{n+1}})\cdots
    \prod^{N_1} C(\l^{C_1})\prod^{M_1} A(\l^{A_1}) \nonumber\\
&\qquad\qquad\qquad
    \prod^{L_1} D(\l^{D_1})\prod^{K_1} B(\l^{B_1})\cdots
	\prod^{L_m} D(\l^{D_m})\prod^{K_m} B(\l^{B_m})\vert 0\rangle\nonumber\\
&\nonumber\\
&=\sum_{A',C'}
\left(\prod\D(\l^{C'},\l^{A'})\right) ^{-1}
\left(\prod\l^{C'}\right)^{2M_{n+1}-1}
\left(\prod\l^{C_{n+1}}\right)\nonumber\\
&\qquad
\langle 0\vert
	\prod^{M_{n+1}} A(\l^{A'}) \prod^{N_{n+1}} C(\l^{C'})
    \prod^{N_n} C(\l^{C_n})\prod^{M_n} A(\l^{A_n})\cdots 
	\prod^{N_1} C(\l^{C_1})\prod^{M_1} A(\l^{A_1})\nonumber\\
&\qquad\qquad\qquad
    \prod^{L_1} D(\l^{D_1})\prod^{K_1} B(\l^{B_1})\cdots
	\prod^{L_m} D(\l^{D_m})\prod^{K_m} B(\l^{B_m})\vert 0\rangle\nonumber\\
&\nonumber\\
&=\prod_{i<j}g(\l_i^{B\vee D},\l_j^{B\vee D}) 
	\prod_{i>j}g(\l_i^{A\vee C},\l_j^{A\vee C})
	J_m\left(\{\l^D\};\{\l^B\}\right)
	(-1)^{LM'} \nonumber\\
&\qquad\qquad\qquad
	\sum_{A',C'} \mathrm{det} G(\{\l^{A'^c}\};\{\l^{B\vee D}\})\prod a(\l^{A'})\times X 
\label{tenkai}
\end{align}
where $\{\l_i^{A'^c}\}_{i=1}^{N+M'}=
\{\l_i^{A\vee C}\}_{i=1}^{N+M}\backslash\{\l_i^{A'}\}_{i=1}^{M_{n+1}}$, 
and the summation is taken over all decompositions 
$\{\l^{A'}\}\cup\{\l^{ C'}\}=\{\l^{A_{n+1}}\}\cup\{\l^{C_{n+1}}\}$. 
Here we used $\langle 0|A(u)=\langle 0|a(u)$. 
We now calculate $X$,  
\begin{align*}
X&
=\left(\prod\D(\l^{C'},\l^{A'})\right) ^{-1}
\left(\prod\l^{C'}\right)^{2M_{n+1}-1}
\left(\prod\l^{C_{n+1}}\right)\\
&\qquad\times
\prod_{i>j} g(\l_i^{A'^c},\l_j^{A'^c})
\left(\prod_{i>j} g(\l_i^{A\vee C},\l_j^{A\vee C})\right)^{-1}\\
&\qquad\times
I_n\left(\{\l^{C_1}\},\cdots,\{\l^{C_n}\}\cup\{\l^{ C'}\};
\{\l^{A_1}\},\cdots,\{\l^{A_n}\}\right)\\
&
=\prod_{i<j} \D(\l_i^{A\vee C},\l_j^{A\vee C})
\left(\D(\l^{C'},\l^{A'})\prod_{i<j}
\D(\l_i^{A'^c},\l_j^{A'^c})\right)^{-1}\\
&\qquad\times
I_n\left(\{\l^{C_1}\},\cdots,\{\l^{C_n}\}\cup\{\l^{ C'}\};
\{\l^{A_1}\},\cdots,\{\l^{A_n}\}\right)\\
&\qquad\times
\left(\prod\l^{A'^c}\right)^{N'+N_{n+1}+M'-1}
	\left(\prod\l^A\prod\l^C\right)^{-(N'+N_{n+1}+M'+M_{n+1}-1)}\\
&\qquad\times
	\left(\prod\l^{C'}\right)^{2M_{n+1}-1}
\left(\prod\l^{C_{n+1}}\right) .
\end{align*}
We deal with the last expression in detail. 
First, 
\begin{align*}
&\prod_{i<j} \D(\l_i^{A\vee C},\l_j^{A\vee C})
\left(\prod\D(\l^{C'},\l^{A'})\prod_{i<j}
\D(\l_i^{A'^c},\l_j^{A'^c})\right)^{-1}\\
&
=(-1)^\sigma
\prod_{i<j} \D(\l_i^{A'},\l_j^{A'})
\prod \D(\l^{A'^c},\l^{A'})
\left( \prod\D(\l^{C'},\l^{A'})\right)^{-1}\\
&
=(-1)^\sigma
\prod_{i<j} \D(\l_i^{A'},\l_j^{A'})
\prod_{i=1}^n \left[\prod \D(\l^{C_i},\l^{A'})\prod \D(\l^{A_i},\l^{A'})\right]\\
&
=(-1)^\sigma
\left(\prod \l^{A'}\right)^{N'+N_{n+1}+2(M'+M_{n+1})-2}\\
&\qquad
\mathrm{det}
\left[
\prod_{i=1}^n \left[\prod \D(\l^{C_i},\l_k^{A'})\prod \D(\l^{A_i},\l_k^{A'})
\right]
\left(\l_k^{A'}\right)^{-(N'+N_{n+1}+2M'+2(\ell-1))}
\right]_{1\le k,\ell\le M_{n+1}}.
\end{align*}
In the last line, we used the Vandermonde determinant. 
Here, $\sigma$ reorders $\{\l^{A\vee C}\}$ into the sequence 
$\{\l^{C'}\}$ followed by $\{\l^{C_n}\},$ $\{\l^{A_n}\},$ $\cdots,$
$\{\l^{A_1}\},$ $\{\l^{A'}\}$.
Next, 
\begin{align*}
&I_n\left(\{\l^{C_1}\},\cdots,\{\l^{C_n}\}\cup\{\l^{ C'}\};
\{\l^{A_1}\},\cdots,\{\l^{A_n}\}\right)\\
&\times
	\left(\prod\l^{A'^c}\right)^{N'+N_{n+1}+M'-1}
	\left(\prod\l^A\prod\l^C\right)^{-(N'+N_{n+1}+M'+M_{n+1}-1)}
	\left(\prod\l^{C'}\right)^{2M_{n+1}-1}\\
&\times
	\left(\prod\l^{C_{n+1}}\right)
	\left(\prod \l^{A'}\right)^{N'+N_{n+1}+2(M'+M_{n+1})-2}\\
&\\
&=\prod_{i=1}^n \left[ 
		\left(\prod\l^{C_i}\right)^{\sum_{j=1}^i M_j-\sum_{j=i+1}^n M_j}
		\left(\prod\l^{A_i}\right)^{\sum_{j=1}^i M_j-\sum_{j=i+1}^n M_j-1}
	\right]
	\left(\prod \l^{C'}\right)^{M'}\\
&\times
	\left(\prod \l^{A} \prod \l^{C} 
	\left(\prod \l^{A'}\right)^{-1}\right)^{N'+N_{n+1}+M'-1}
	\left(\prod\l^A\prod\l^C\right)^{-(N'+N_{n+1}+M'+M_{n+1}-1)}\\
&\times
	\left(\prod\l^{C'}\right)^{2M_{n+1}-1}
	\left(\prod\l^{C_{n+1}}\right)
	\left(\prod \l^{A'}\right)^{N'+N_{n+1}+2(M'+M_{n+1})-2}\\
%&\\
\end{align*}
\begin{align*}
&=\prod_{i=1}^n \left[ 
		\left(\prod\l^{C_i}\right)^{\sum_{j=1}^i M_j-\sum_{j=i+1}^n M_j}
		\left(\prod\l^{A_i}\right)^{\sum_{j=1}^i M_j-\sum_{j=i+1}^n M_j-1}
	\right]\\
&\times
	\prod_{i=1}^n \left[ 
	\prod \l^{C_i}\prod \l^{A_i}
	\right]^{-M_{n+1}}
	\left(\prod \l^{C'} \prod \l^{A'}\right)^{M'+M_{n+1}-1}
	\left(\prod\l^{C_{n+1}}\right)\\
&\\
&=\prod_{i=1}^n \left[ 
		\left(\prod\l^{C_i}\right)^{\sum_{j=1}^i M_j-\sum_{j=i+1}^{n+1} M_j}
		\left(\prod\l^{A_i}\right)^{\sum_{j=1}^i M_j-\sum_{j=i+1}^{n+1} M_j-1}
	\right]\\
&\times
	\left( \prod \l^{C_{n+1}}\right)^{M'+M_{n+1}} 
	\left( \prod \l^{A_{n+1}}\right)^{M'+M_{n+1}-1}\\
&\\
&=\prod_{i=1}^{n+1} \left[ 
		\left(\prod\l^{C_i}\right)^{\sum_{j=1}^i M_j-\sum_{j=i+1}^{n+1} M_j}
		\left(\prod\l^{A_i}\right)^{\sum_{j=1}^i M_j-\sum_{j=i+1}^{n+1} M_j-1}
	\right]\\
&\\
&
=I_{n+1}\left(\{\l^{C_1}\},\cdots,\{\l^{C_{n+1}}\};
\{\l^{A_1}\},\cdots,\{\l^{A_{n+1}}\}\right).
\end{align*}
Note that 
$\prod \l^{A'} \prod \l^{C'} =\prod \l^{A_{n+1}} \prod \l^{C_{n+1}}$. 
Going back to (\ref{tenkai}), substituting the above calculations, we get
\begin{align*}
&\prod_{i<j}g(\l_i^B,\l_j^B) \prod_{i>j}g(\l_i^{A\vee C},\l_j^{A\vee C})
I_{n+1}\left(\{\l^{C_1}\},\cdots,\{\l^{C_{n+1}}\};
\{\l^{A_1}\},\cdots,\{\l^{A_{n+1}}\}\right)\\
&\qquad J_m\left(\{\l^D\};\{\l^B\}\right) (-1)^{LM'}
\sum_{A',C'} 
(-1)^\sigma \mathrm{det}\ G\left(\{\l^{A'^c}\};\{\l^{B\vee D}\}\right)
\prod a(\l^{A'})\\
&\qquad
\mathrm{det}
\left[
\prod_{i=1}^n \left[\prod \D(\l^{C_i},\l_k^{A'})\prod \D(\l^{A_i},\l_k^{A'})\right]
\left(\l_k^{A'}\right)^{-(N'+N_{n+1}+2M'+2(\ell-1))}
\right]_{1\le k,\ell\le M_{n+1}}\\
&\\
&
=\prod_{i<j}g(\l_i^B,\l_j^B) \prod_{i>j}g(\l_i^{A\vee C},\l_j^{A\vee C})
I_{n+1}\left(\{\l^{C}\};\{\l^{A}\}\right)
J_m\left(\{\l^D\};\{\l^B\}\right) (-1)^{LM'}\\
&\qquad\sum_{A',C'} 
(-1)^{\sigma'+LM_{n+1}} \mathrm{det}\ G\left(\{\l^{A'^c}\};\{\l^{B\vee D}\}
\right)
\mathrm{det}
\left[ h_{n+1,\ell}(\l_k^{A'})\right]
_{1\le k\le M_{n+1} \atop N+M'+1\le \ell\le N+M'+M_{n+1}}
\end{align*}
where $\sigma'$ sends the indices of row with $\{\l^{A'}\}$ for 
matrix $G\left(\{\l^{A\vee C}\};\{\l^{B\vee D}\}\right)$ to the last. 
There appears another sign $(-1)^{LM_{n+1}}$ 
because $L$ lines of the matrix $G$ with $1\le a\le m$ are passed through 
by $M_{n+1}$ lines. 
We can extend the summation of $A', C'$ to all 
$\{\l^{A'}\}\subset\{\l^{A\vee C}\}$ since 
$\mathrm{det}
\left[ h_{n+1,\ell}(\l_k^{A'})\right] =0$
unless $\{\l^{A'}\}\subset\{\l^{A_{n+1}}\}\cup\{\l^{C_{n+1}}\}$. 
As a result, the sum becomes 
\[
(-1)^{LM_{n+1}} \mathrm{det}\ G\left(\{\l^{A\vee C}\};\{\l^{B\vee D}\}\right).
\] 
The whole output is exactly the $n+1$ case of the formula. 
Computation for $m+1$ case is done in the same way by developing relations 
(\ref{DB1}) and (\ref{DB2}). 
Then by induction the proof is completed.
\end{proof}

\section{Generating Function of Boxed Plane Partition}\label{genefuncbpp}

As an application of the formula in the previous section, 
we evaluate the determinant of the generalized scalar product corresponding to
the generating function of the boxed plane partition 
in an $N\times L\times M$-box. 
\begin{proposition}
The following product expression holds for the generalized scalar product. 
\begin{align}
&\langle 0\vert \prod_{k=L-N+1}^L C(q^{-k/2}) \prod_{k=1}^{L-N} A(q^{-k/2})
\prod_{k=1}^N B(q^{(k-1)/2}) \vert 0\rangle \nonumber\\
&\qquad=
q^{{-\frac{1}{2}LMN+\frac{1}{4}(L-N)(M+1)(L-N+1)}}
\prod_{i=1}^L \prod_{j=1}^N
\frac{1-q^{i+j+M-1}}{1-q^{i+j-1}}
\end{align}
\end{proposition}

\begin{proof}
For an evaluation of the relevant determinant in (\ref{detformula}), 
we use the following determinant formula extending the one in \cite{Ku}: 
\begin{align}
\mathrm{det}
\Big[
&\left( 
\frac{s^{(i+j-1)/2}-s^{-(i+j-1)/2}}{q^{(i+j-1)/2}-q^{-(i+j-1)/2}} 
\right)_{1\le i\le L \atop 1\le j\le N}
\left( \left(sq^{2j-1}\right)^{i/2}\right)_{1\le i\le L \atop 1\le j\le L-N} 
\Big] \nonumber\\
&=(-1)^{\frac{1}{2} N(N-1)}
q^{\frac{1}{4} (L-N)(L^2+N^2-LN+L-N)} s^{\frac{1}{4} (L-N)(L-N+1)}
(q^{1/2}-q^{-1/2})^{\frac{1}{2} L(L-1)+\frac{1}{2} N(N-1)} \nonumber\\
&\qquad\times
	\prod_{k=0}^{N-1} [\sigma-k]^{N-k} \prod_{k=1}^{L-N} [\sigma+k]^N
	\prod_{k=L-N+1}^{L-1} [\sigma+k]^{L-k} \nonumber\\
&\qquad\times
	\prod_{1\le j<i\le N} [i-j]^2 \prod_{N+1\le j<i\le L} [i-j]
	\prod_{L-N+1\le i\le L \atop 1\le j\le N} [i+j-1]^{-1}
\label{genKu}
\end{align}
where we put $s=q^\sigma=q^{M+N}$ and $[k]=\frac{q^{k/2}-q^{-k/2}}{q^{1/2}-q^{-1/2}}$. 
The remaining computation for the proof is a straightforward exercise. 

Eq.(\ref{genKu}) is briefly proved as follows. 
The determinant of the matrix multiplied by 
$s^{i/2}$ for rows with $1\le i\le L$ and 
$s^{(j-1)/2}$ for columns with $1\le j\le N$ 
is a polynomial in $s$ of degree $LN+(L-N)(L-N+1)/2$. 
If we put $s=q^k,\ 0\le k\le N-1$, the matrix is spanned by linearly independent 
vectors $\mbox{\bfseries\itshape{x}}_j=\left( q^{(2j+1-k)i/2}\right),\ 1\le j\le k$,
and $\mbox{\bfseries\itshape{y}}_j=\left( q^{(2j+1+k)i/2}\right),\ 1\le j\le L-N$,
and its rank is $k+L-N$, so we have the divisor $(s-q^k)^{N-k}$. 
If we put $s=q^{-k},\ 1\le k\le L-1$, the matrix is spanned by linearly 
independent vectors 
$\mbox{\bfseries\itshape{x}}_j=
\left( q^{(2j+1-k)i/2}\right),\ 1\le j\le \mathrm{max}\{k,\ L-N\}$, 
and its rank is $\mathrm{max}\{k,\ L-N\}$, so we 
have the divisor $(s-q^{-k})^N$ for $1\le k\le L-N$ and $(s-q^{-k})^{L-k}$ for
$L-N+1\le k\le L-1$. We only need the overall factor. It is obtained 
by examining the coefficient of the lowest degree term of $s$. It turns out 
to be the product of a Cauchy-type determinant and a Vandermonde-type determinant, 
so it can be easily calculated. 
\end{proof}

\section{Conclusion}
\label{concl}

In this paper, we have introduced and calculated a generalized scalar product 
of the integrable phase model. 
We have shown  in two ways that a generalized scalar product with generic spectral parameters 
is expressed as a determinant.
First, we have shown that the matrix elements of the monodromy operators are 
expressed in terms of the skew Schur functions through graphical interpretations; 
vertical lattices with arrows and lattice paths on them. 
A generalized scalar product is expressed as the sum of products of the skew 
Schur functions, where the sum is taken over all the partitions satisfying a restriction 
$\lambda_{1}\le M$.  The sum is reexpressed as a determinant, applying the 
Cauchy-Binet formula. 
Second, the commutation relations for the monodromy operators of the QISM allow 
us to express a generalized scalar product as a determinant of matrix with elements 
of two-point functions and one-point functions. 
This expression is also valid for the inhomogeneous phase model.  

We have found that a generalized scalar product is related to 
the generating function of the skew plane partition. 
More precisely, a generalized scalar product with the special choice of the spectral 
parameters is up to some factor equal to the generating function of 
the corresponding skew plane partition. 
A sequence of the monodromy operators determines the skew shape. 

The skew plane partition in the semi-infinite box was considered 
in terms of the free fermion representation in~\cite{OkRe1, OkRe2}. 
The generating function of the plane partition 
can be written as the expectation value of products of successive vertex operators. 
The phase model representation here of the skew plane partition has some advantages 
when restricted in a box. In the fermion formalism this restriction requires us 
to insert projection operators between vertex operators and 
makes it difficult to calculate the generating function explicitly. 
Correlation functions are easily treated in the fermion formalism. 
To find an explicit expression of general correlation functions in the phase model 
formalism is remained as a future problem.

\section*{Acknowledgement}
The authors would like to thank Professor Miki Wadati for critical reading of the 
manuscript and giving useful comments.

\end{document}